\documentclass[aps,prd,preprintnumbers,superscriptaddress,nofootinbib,showpacs]{revtex4}%
\usepackage[dvips]{graphicx}
\usepackage{here}
\usepackage{bm,latexsym,amsmath,amssymb,amsfonts,mathrsfs}
\usepackage{color}
\input{colordvi.tex}
\usepackage[dvipdfm]{hyperref}
\usepackage{url}
\hypersetup{
    colorlinks=true,
    citecolor=cyan,
}
\newcommand*{\D}{\textrm{d}}
\newcommand*{\mpl}{M_\textrm{Pl}}
\begin{document}

\title{Primordial non-Gaussianities of gravitational waves beyond Horndeski}

\author{Yuji~Akita}
\email[Email: ]{y.akita"at"rikkyo.ac.jp}
\affiliation{Department of Physics, Rikkyo University, Toshima, Tokyo 175-8501, Japan
}

\author{Tsutomu~Kobayashi}
\email[Email: ]{tsutomu"at"rikkyo.ac.jp}
\affiliation{Department of Physics, Rikkyo University, Toshima, Tokyo 175-8501, Japan
}

\preprint{RUP-15-28}

\begin{abstract}

We clarify the features of primordial non-Gaussianities of tensor perturbations in Gao's
unifying framework of scalar-tensor theories. The general Lagrangian is given in terms of the
ADM variables so that the framework maintains spatial covariance
and includes the Horndeski theory and Gleyzes-Langlois-Piazza-Vernizzi (GLPV) generalization
as specific cases.
It is shown that the GLPV generalization does not give rise to
any new terms in the cubic action compared to the case of the Horndeski theory,
but four new terms appear in more general theories beyond GLPV.
We compute the tensor 3-point correlation functions analytically
by treating the modification to the dispersion relation as a perturbation.
The relative change in the 3-point functions
due to the modified dispersion relation is only mildly configuration-dependent.
When the effect of the modified dispersion relation is small,
there is only a single cubic term generating squeezed non-Gaussianity,
which is the only term present in general relativity.
The corresponding non-Gaussian amplitude has a fixed and universal feature,
and hence offers a ``consistency relation'' for primordial tensor modes
in a quite wide class of single-field inflation models.
All the other cubic interactions are found to give peaks at equilateral shapes.
\end{abstract}

\maketitle

\section{Introduction}

Inflation~\cite{Guth:1980zm,Sato:1980yn,Starobinsky:1980te},
the accelerated expansion of the early universe, is an almost perfect idea
for generating seeds for structure formation as well as resolving several issues in
standard Big Bang cosmology.
Valuable information about the physics of inflation is carried by
the power spectrum and bispectrum of primordial perturbations.
The properties of the primordial curvature perturbations, ${\cal R}$,
can be explored through observations of CMB anisotropies~\cite{Ade:2015lrj}
and are characterized primarily by the amplitude and the spectral index,
which can be translated to information about the shape of the inflaton potential.
Some nonstandard inflation models predict strongly non-Gaussian curvature perturbations~\cite{Chen:2006nt},
and hence can be constrained by the bispectrum of ${\cal R}$~\cite{Ade:2015ava}.
In addition to the curvature perturbations, tensor modes, {\em i.e.},
gravitational waves, are also produced during the phase of the inflationary expansion.
Currently, we have only an upper bound on the tensor-to-scalar ratio $r$,
but primordial gravitational waves could be a smoking gun of inflation if observed in future experiments
of direct detection or CMB B-mode polarization measurements.

In this paper, we study primordial non-Gaussianity of gravitational waves
from inflation.
After the seminal work by Maldacena who computed not only
the scalar 3-point correlation function but also
the tensor 3-point function~\cite{Maldacena:2002vr},
several authors have investigated non-Gaussian signatures of primordial tensor
modes~\cite{Maldacena:2011nz,Soda:2011am,McFadden:2011,Gao:2011vs,Gao:2012ib,Huang:2013epa,Bzowski:2011ab,Zhu:2013fja,Cook:2013xea,Sreenath:2013xra,Noumi:2014zqa,Sreenath:2014nka,Fu:2015vja,Chowdhury:2015cma}.
As tensor non-Gaussianity could in principle be measured {\em e.g.}, via the bispectrum of
B-mode fluctuations, it offers us yet another discriminant among an enormous number of different
inflation models. The purpose of the present paper is therefore
to clarify the features of primordial tensor non-Gaussianities
within a framework involving as many inflation models as possible.

Generalized G-inflation~\cite{Kobayashi:2011nu}
is the general framework to study single-field inflation models
based on the Horndeski theory~\cite{Horndeski:1974wa,Deffayet:2011gz},
which is the most general scalar-tensor theory with
second-order field equations and thus is free from Ostrogradsky ghost instabilities.
Within this generalized G-inflation framework,
the general form of the power spectra of curvature and tensor perturbations
has been obtained in Ref.~\cite{Kobayashi:2011nu}, and
the cubic interactions of ${\cal R}$
have been derived in Refs.~\cite{Gao:2011qe,DeFelice:2011uc,RenauxPetel:2011sb,Ribeiro:2011ax,DeFelice:2013ar}
to evaluate primordial non-Gaussianity of the curvature perturbations.
Tensor 3-point interactions in the Horndeski theory have been classified completely in Ref.~\cite{Gao:2011vs}
and it has been shown that there are only two independent contributions:
the ``standard'' one that is present already in general relativity and generates
squeezed non-Gaussianity, and the other ``nonstandard'' one 
predicting equilateral non-Gaussianity that
arises from
the coupling between the Einstein tensor and derivatives of the scalar field.
In particular, the former contribution has the fixed non-Gaussian amplitude
irrespective of an underlying model.

Recently, it was noticed that the Horndeski theory can further be generalized
to higher derivative theories that nevertheless preserve the same propagating degrees of freedom,
{\em i.e.}, two polarizations of gravitational waves and one scalar, and hence
remain free of the Ostrogradsky ghost,
giving Gleyzes-Langlois-Piazza-Vernizzi (GLPV) generalization of the Horndeski
theory~\cite{Gleyzes:2014dya,Gleyzes:2014qga}.
The GLPV theory was then extended to a more general, unifying framework of
scalar-tensor theory by Gao~\cite{Gao:2014soa}.
Linear cosmological perturbations in this general class of models have been
investigated in
Refs.~\cite{Gleyzes:2014dya,Gleyzes:2014qga,Gao:2014soa,Kase:2014cwa,DeFelice:2014bma,DeFelice:2015isa,Fujita:2015ymn}
and non-Gaussianity of the curvature perturbations in Ref.~\cite{Fasiello:2014aqa}.
In this paper, we determine the non-Gaussian features
of primordial tensor perturbations in Gao's unifying framework of general scalar-tensor theories.

This paper is organized as follows. 
In the next section, we give a brief review of Gao's unifying framework and extract the terms relevant
to tensor perturbations out of all the possible terms.
In Sec.~III, the linear theory of the tensor perturbations
is presented to give the primordial power spectrum.
Then we move to the cubic action and compute the 3-point correlation functions of tensor modes
in Sec.~IV. 
We put rough constraints on non-Gaussianities of tensor modes from
CMB observations in Sec.~V.
Finally, Sec.~VI is devoted to discussions and conclusions.

\section{General Lagrangian for tensor perturbations}

Let us begin with a brief review on the unifying framework of scalar-tensor theories.
The most general scalar-tensor theory with second-order field equations
is given by the Horndeski Lagrangian~\cite{Horndeski:1974wa}. As shown in Ref.~\cite{Kobayashi:2011nu},
it turns out that the Horndeski Lagrangian can be written equivalently in the modern form of
the generalized Galileon~\cite{Deffayet:2011gz} as
\begin{align}
\frac{{\cal L}_H}{\sqrt{-g}}=
G_2(\phi, X)-G_3(\phi,X)\Box\phi+G_4(\phi, X)R^{(4)}+
G_5(\phi, X)G_{\mu\nu}^{(4)}\nabla^\mu\nabla^\nu\phi\cdots,\label{Hor-Cov}
\end{align}
where $R^{(4)}$ and $G_{\mu\nu}^{(4)}$ are the (four-dimensional) Ricci scalar and
the Einstein tensor, respectively, and
we have four arbitrary functions of $\phi$ and $X:=-g^{\mu\nu}\partial_\mu\phi\partial_\nu\phi/2$.
Performing an ADM decomposition by taking $\phi=\;$const hypersurfaces as constant time hypersurfaces,
one obtains the Lagrangian of the form
\begin{align}
\frac{{\cal L}_H}{\sqrt{-g}}&=A_2(t,N)+A_3(t,N)K+
A_4(t,N)\left(K^2-K_{ij}K^{ij}\right)+B_4(t,N)R
\notag \\ &
\quad +A_5(t,N)\left(K^3-3KK_{ij}K^{ij}+2K_{ij}K^{jk}K_{k}^{i}\right)
+B_5(t,N)K^{ij}\left(R_{ij}-\frac{1}{2}g_{ij}R\right),\label{Hor-ADM}
\end{align}
where $K_{ij}$ and $R_{ij}$ are the extrinsic and intrinsic curvature tensors
on the constant $\phi$ hypersurfaces.
The functions of $\phi$ and $X\;(=\dot\phi^2/2N^2)$ in the covariant Lagrangian~(\ref{Hor-Cov})
are now regarded as the functions of the time coordinate $t$ and the Lapse function $N$
in the ADM language.
We thus should have four functions of $t$ and $N$, and indeed
$A_4$, $A_5$, $B_4$, and $B_5$ obey
\begin{align}
A_4=-B_4-N\frac{\partial B_4}{\partial N},
\quad
A_5=\frac{N}{6}\frac{\partial B_5}{\partial N},\label{Hor-rel}
\end{align}
leaving four arbitrary functions as expected.
This is the unitary gauge description of the Horndeski theory
which is particularly useful in the study of cosmology.

The scalar degree of freedom is apparently hidden in the unitary gauge description.
However, variation of~(\ref{Hor-ADM}) with respect to $N$ yields a second-class constraint
eliminating one degree of freedom rather than two.
This fact signifies the presence of the scalar degree of freedom.
One notices here that the
relations~(\ref{Hor-rel}) play no role when counting the degrees of freedom.
Therefore, the degrees of freedom remain the same even if one liberates $A_4$ and $A_5$
from~(\ref{Hor-rel}) and considers the theory with six arbitrary functions of $t$ and $N$.
This trick was first used by Gleyzes {\it et al.}~\cite{Gleyzes:2014dya} to
construct the GLPV scalar-tensor theory beyond Horndeski.
Since the GLPV theory is more general than Horndeski,
the covariant equations of motion are of higher order in general. Nevertheless,
the true propagating degrees of freedom are one scalar and two tensors
as in the Horndeski theory~\cite{Gleyzes:2014dya,Lin:2014jga,Gleyzes:2014qga}.
(Note, however, that counting the degrees of freedom using the unitary gauge Lagrangian
involves subtle points~\cite{Deffayet:2015qwa,Langlois:2015cwa}.)
See also Refs.~\cite{Zumalacarregui:2013pma,Ohashi:2015fma} for attempts to extend the Horndeski theory in different
directions.

The GLPV theory can further be generalized while retaining the number of propagating degrees of freedom
as demonstrated by Gao~\cite{Gao:2014soa,Gao:2014fra}.
The key is to notice that one no longer needs to impose the specific
combinations of $K_{ij}$ and $R_{ij}$ as presented in Eq.~(\ref{Hor-ADM}).
Now one arrives at the unifying framework of scalar-tensor theories
which is given by the sum of three dimensional scalars composed of $K_{ij}$, $R_{ij}$, and
$a_i:=\partial_i\ln N$, with coefficients depending on $t$ and $N$:
\begin{align}
\frac{{\cal L}}{\sqrt{-g}}&=
d_0(t, N)+d_1(t, N)R+d_2R^2+\cdots+d_4a_ia^i+\cdots+\left(a_0+a_1R+\cdots\right)K
\notag \\ & \quad
+\left(a_2R^{ij}+\cdots\right)K_{ij}+b_1K^2+b_2K_{ij}K^{ij}+\cdots
+c_1K^3+c_2KK_{ij}K^{ij}+c_3K_{ij}K^{jk}K_{k}^{i}+\cdots.
\end{align}
The ``six-parameter'' subclass of the above Lagrangian with
\begin{align}
& a_0=A_3, \quad  -2a_1=a_2=B_5, \quad b_1=-b_2=A_4,
\notag\\
&c_1=-c_2/3=c_3/2=A_5, \quad d_0=A_2, \quad d_1=B_4,
\notag\\
&{\rm others}\;=0,
\end{align}
corresponds to
the GLPV theory.
By imposing the further restrictions~(\ref{Hor-rel})
we have the ``four-parameter'' subclass which reproduces the Horndeski theory.
Even (the healthy extension of) Ho\v{r}va gravity~\cite{Horava:2009uw,Blas:2009qj}
is included within this framework.
Among possible terms Gao considers 28 operators satisfying the requirements that
(i) there are no derivatives higher than two when going back to the covariant Lagrangian
and (ii) the number of second-order derivative operators does not exceed three~\cite{Gao:2014soa}.
We shall work in Gao's framework to cover a fairly wide range of single-field inflation models,
and evaluate all possible tensor 3-point functions.
As explained below, the number of operators that are relevant to
tensor perturbations is in fact not as large as 28.

Let us now simplify the Lagrangian before computing
the quadratic and cubic terms of tensor perturbations.
We define the tensor perturbations $h_{ij}$ by
\begin{align}
\D s^2&=-\D t^2+\gamma_{ij}\D x^i\D x^j
\notag \\
&= -\D t^2+a^2\left(e^h\right)_{ij}\D x^i\D x^j,
\end{align}
where $(e^h)_{ij}=\delta_{ij}+h_{ij}+(1/2)h_{ik}h_{kj}+\cdots$
and $h_{ij}$ is transverse and traceless.
We simply have $\sqrt{-g}=a^3$ with this definition.
As a consequence of
the traceless and transverse nature of $h_{ij}$, the trace of the extrinsic curvature tensor,
$K\sim \gamma^{ij} \partial_t \gamma_{ij}$, can be written solely in terms of the background quantities.
One sees that the intrinsic curvature scalar $R$ does not contain any first-order terms
up to total derivatives. These facts lead us to
the following 14 terms that are relevant to the tensor perturbations:
\begin{eqnarray}
	\frac{{\cal L}}{\sqrt{-g}}&=&\left(a_1R+a_4R_i^jR_j^i \right)K
	+\left(a_2R_i^j+a_7R_i^kR_k^j \right)K_j^i
	+b_3RK^2+\left(b_2+b_4R \right)K_i^jK_j^i
		\nonumber\\&&
	+\left(b_5KK_i^j+b_6K_i^kK_k^j \right)R_j^i
	+c_2KK_i^jK_j^i+c_3K_i^jK_j^kK_k^i
	 + d_1R+d_3R_i^jR_j^i+d_7R_i^jR_j^kR_k^i.\label{effL1}
\end{eqnarray}
Here, the coefficients may be regarded as the functions of $t$ since we have set $N=1$.
It is convenient to split the extrinsic curvature into the background and perturbation parts as
\begin{eqnarray}
	K_i^j=H\delta_i^j +\delta K_i^j,\label{k-split}
\end{eqnarray}
where $H:=\dot a/a$ and
\begin{eqnarray}
	\delta K_i^j=\frac{1}{2}\dot h_{ij}+\frac{1}{4}\left(h_{ik}\dot h_{kj}-\dot h_{ik}h_{kj}\right)
	+\mathcal O(h^3),
\end{eqnarray}
with a dot standing for differentiation with respect to $t$.
It can be seen directly that $\delta K_i^j$ is traceless, $\delta K_i^i=0$.
Substituting Eq.~(\ref{k-split}) to the Lagrangian~(\ref{effL1})
and extracting the perturbations,
we obtain
\begin{eqnarray}
	\frac{{\cal L}}{\sqrt{-g}}&=&
	\widetilde d_1 R+\widetilde d_3 R_i^jR_j^i + d_7 R_i^jR_j^kR_k^i
	+\widetilde b_2\delta K_i^j\delta K_j^i + c_3\delta K_i^j\delta K_j^k\delta K_k^i
		\nonumber\\&&
	+\widetilde a_2R_i^j\delta K_j^i
	+a_7R_i^jR_j^k\delta K_k^i
	+b_6R_i^j\delta K_j^k\delta K_k^i, \label{eq:relevant action}
\end{eqnarray}
where the coefficients are defined as
\begin{eqnarray}
	\widetilde d_1&=&d_1+\left(3a_1+a_2\right)H+\left(9b_3+3b_4+3b_5+b_6 \right)H^2,
	\\
	\widetilde d_3&=&d_3+\left(3a_4+a_7 \right)H,
	\\
	\widetilde b_2&=& b_2+3\left(c_2+c_3 \right)H,
	\\
	\widetilde a_2&=&a_2+\left(3b_5+2b_6 \right)H.
\end{eqnarray}
Those coefficients are determined
once the concrete form of the Lagrangian is fixed and the background cosmological evolution is obtained.
The reduced
Lagrangian~(\ref{eq:relevant action}) is sufficient for deriving the most general quadratic and cubic
Lagrangians for the tensor perturbations
within Gao's unifying framework of scalar-tensor theories.

In the GLPV subclass, we have only four non-vanishing coefficients,
$\widetilde d_1$, $\widetilde b_2$, $c_3$, and $\widetilde a_2$, all of which are arbitrary.
In the Horndeski case, $\widetilde b_2$ and $c_3$ are fixed by $\widetilde d_1$ and $\widetilde a_2$
via the relations~(\ref{Hor-rel}), and hence we have essentially two functional degrees of freedom,
though all four of these coefficients are still non-vanishing.
Thus, even if one generalizes the Horndeski theory to GLPV,
no new terms appear in the Lagrangian for the tensor perturbations,
though the coefficients can be chosen more freely in the GLPV theory.
The other four terms appear for the first time when going to Gao's framework.
In particular, the $a_7$ and $b_6$ terms are completely new, while
the $\widetilde d_3$ and $d_7$ terms can be found in the specific case of Ho\v{r}ava gravity as well.

\section{Primordial power spectrum with modified dispersion relation}

\subsection{Linear theory}

From Eq.~(\ref{eq:relevant action}) we obtain the quadratic action
for the tensor perturbations~\cite{Gao:2014soa,DeFelice:2014bma,Fujita:2015ymn}:
\begin{eqnarray}
	S&=&\frac{1}{8}\int\D t\D^3x\;a^3\left[ \mathcal{G}_T\dot h_{ij}^2
	-\frac{\mathcal{F}_T}{a^2}\left(\partial_k h_{ij}\right)^2
	+2\frac{\tilde d_3}{a^4}\left(\partial^2 h_{ij}\right)^2\right],\label{quadaction}
\end{eqnarray}
where
\begin{eqnarray}
	\mathcal{G}_T:=2\widetilde{b}_2,
	\quad
	\mathcal{F}_T:=2\widetilde{d_1}+\dot{\widetilde{a}}_2+H \widetilde{a}_2.
\end{eqnarray}
The third term modifies the dispersion relation.
This term is absent in the Horndeski theory and even in its GLPV generalization,
but it appears
in more general theories in the unifying framework.
The goal of this section is to present the power spectrum of primordial tensor modes
with the modified dispersion relation.
Primordial perturbations with this type of the dispersion relation
have been studied {\em e.g.}, in Refs.~\cite{Martin:2002kt,Ashoorioon:2011eg,Kobayashi:2015gga},
with an emphasis on the curvature perturbation.

Before calculating the linear solution and the power spectrum,
let us comment on the stability conditions derived from the quadratic action~(\ref{quadaction}).
It is required that $\mathcal{G}_T>0$ to avoid ghost instabilities.
We also impose the condition
\begin{eqnarray}
\widetilde d_3\le 0
\end{eqnarray}
in order for the modes with high momenta to be stable.
If ${\cal F}_T>0$ then the tensor perturbations are stable at low momenta as well.
(However, in fact
${\cal F}_T$ can be negative for a short period provided that $\widetilde d_3<0$,
because the instability grows only at low momenta and hence is not catastrophic.)

We move to the Fourier space,
\begin{eqnarray}
	h_{ij}(t,{\bf x})=\int \frac{\D^3k}{(2\pi)^3} \tilde h_{ij}(t,{\bf k})e^{i\mathbf{k}\cdot\mathbf{x}},
\end{eqnarray}
and solve the equation of motion for $\widetilde h_{ij}(t,\mathbf{k})$,
\begin{align}
\frac{1}{a^2}\frac{\D}{\D\eta}\left(a^2{\cal G}_T\frac{\D}{\D\eta}\widetilde h_{ij}\right)
+\left( {\cal F}_Tk^2-\frac{2\widetilde d_3}{a^2}k^4\right)\widetilde h_{ij}=0,\label{perteq}
\end{align}
where we have introduced the conformal time defined by $\D\eta = \D t/a$.

To obtain an exact solution to Eq.~(\ref{perteq}) and
manipulate the non-Gaussianity in a tractable way,
we make the following assumptions on the background quantities.
First, the background cosmological evolution is assumed to be exact de Sitter,
so that the scale factor is given by $a=1/H(-\eta)$.
Second, ${\cal F}_T$, ${\cal G}_T$, $\widetilde d_3$, and also
the other coefficients in the Lagrangian~(\ref{eq:relevant action})
are all assumed to be constant.
The second assumption is natural and consistent in view of the first assumption,
though strictly speaking the coefficients of the terms involving $R_{ij}$,
{\em i.e.}, $\widetilde d_1$, $\widetilde d_3$, $\cdots$, can be dependent on time
as the background evolution is insensitive to those coefficients
and is determined by the terms constructed only from $K_{ij}$.

With the above assumptions, the dispersion relation can be written as
\begin{eqnarray}
\omega^2=c_h^2k^2+\epsilon^2k^4\eta^2,
\end{eqnarray}
where we defined the dimensionless quantities
\begin{eqnarray}
c_h^2:=\frac{{\cal F}_T}{{\cal G}_T},
\quad
\epsilon^2:=-2H^2\frac{\widetilde d_3}{{\cal G}_T}.
\end{eqnarray}
The normalized mode solution to the perturbation equation~(\ref{perteq}) is given
in terms of the Whittaker function $W$ by
\begin{eqnarray}
	\psi_{\mathbf{k}}(\eta)=
	\frac{\sqrt{2}}{a}\frac{e^{-\pi c_h^2/8\epsilon}}{\sqrt{-\mathcal{G}_T\epsilon k^2\eta}}
	W\left(\frac{i c_h^2}{4\epsilon},\frac{3}{4},-i\epsilon k^2 \eta^2\right).\label{ex-Whitt}
\end{eqnarray}

In the case of the standard dispersion relation, $\epsilon \to 0$,
which is the case for the GLPV theory as well as Horndeski,
we have
\begin{eqnarray}
\psi_{\mathbf{k}}\to \frac{\sqrt{\pi}}{a}\sqrt{\frac{-\eta}{{\cal G}_T}}H_{3/2}^{(1)}(-c_hk\eta)
\end{eqnarray}
up to a phase factor, where $H_{3/2}^{(1)}$ is the Hankel function of the first kind.
Thus, the familiar mode function is reproduced.
For $\epsilon k^2\eta^2 \gg 1$ with $\epsilon\neq 0$,
$(a\sqrt{{\cal G}_T}/2)\psi_{\mathbf{k}}$ reduces to the positive frequency WKB solution,
\begin{eqnarray}
\frac{a\sqrt{{\cal G}_T}}{2}\psi_{\mathbf{k}}\approx \frac{1}{\sqrt{2\omega}}
\exp\left[-i\int^\eta\omega(\eta')\D\eta'\right],
\end{eqnarray}
with $\omega \approx -\epsilon k^2\eta$,
showing that $(a\sqrt{{\cal G}_T}/2)\psi_{\mathbf{k}}$
is indeed canonically normalized. In the opposite, superhorizon limit, $-k\eta\to 0$,
we find
\begin{eqnarray}
\psi_{\mathbf{k}}\to \sqrt{\frac{\pi}{2}}
\frac{e^{-\pi c_h^2/8\epsilon}}{\Gamma(5/4-ic_h^2/4\epsilon)}\frac{H}{{\cal G}_T^{1/2}\epsilon^{3/4}k^{3/2}}
\end{eqnarray}
up to a phase factor.

Using the mode function $\psi_{\mathbf{k}}$, the quantized tensor perturbation can be expanded as
\begin{eqnarray}
	\widetilde{h}_{ij}(t,{\bf k})=\sum_s \left[\psi_{\bf k} e_{ij}^{(s)}({\bf k}) a_s({\bf k})
	+\psi^\ast_{-{\bf k}} e_{ij}^{\ast(s)}(-{\bf k}) a_s^\dagger(-{\bf k})\right],
\end{eqnarray}
where
the transverse and traceless
polarization tensor with the helicity states $s=\pm 2$, $e_{ij}^{(s)}({\bf k})$,
is normalized as
$e_{ij}^{(s)}({\bf k})e_{ij}^{\ast(s')}({\bf k})=\delta_{ss'}$
and has the properties such that
$e_{ij}^{\ast(s)}({\bf k})=e_{ij}^{(-s)}({\bf k})=e_{ij}^{(s)}({\bf -k})$.
The
commutation relation is
given by
$[a_s({\bf k}),~a_{s'}^\dagger({\bf k}')]=(2\pi)^3\delta_{ss'}\delta^{(3)}({\bf k}-{\bf k}')$.
Now we are ready to compute the primordial power spectrum.
First we define ${\cal P}_{ij,kl}(\mathbf{k})$ by
\begin{eqnarray}
	\langle \widetilde{h}_{ij}({\bf k}) \widetilde{h}_{kl}({\bf k}') \rangle= (2\pi)^3\delta^{(3)}({\bf k}+{\bf k}') 	
	\mathcal{P}_{ij,kl}({\bf k}).
\end{eqnarray}
We have
\begin{eqnarray}
	\mathcal{P}_{ij,kl}({\bf k})=\left| \psi_{\bf k} \right|^2 \Pi_{ij,kl}({\bf k}),
\end{eqnarray}
with
\begin{eqnarray}
\Pi_{ij,kl}({\bf k}):=\sum_s e_{ij}^{(s)}({\bf k})e_{kl}^{\ast(s)}({\bf k}).
\end{eqnarray}
The power spectrum, $\mathcal P_h:=(k^3/2\pi^2)\mathcal P_{ij,ij}$, is then given by
\begin{eqnarray}
	\mathcal{P}_h(k)
	=2 \frac{H^2}{\pi^2} \frac{\mathcal{G}_T^{1/2}}{\mathcal{F}_T^{3/2}}
	|F(\epsilon/c_h^2)|^2 ,
\end{eqnarray}
where we defined
\begin{eqnarray}
F(x):=\frac{\sqrt\pi}{2} \frac{x^{-3/4}e^{-\pi/(8x)} }{\Gamma(5/4-i/(4x))}.
\end{eqnarray}
Since $|F(x)|=1-5x^2/8+{\cal O}(x^4)$, the above expression includes, as the $x\to 0$ limit,
the general result with the standard dispersion relation
derived from generalized G-inflation~\cite{Kobayashi:2011nu}.

\subsection{Small $\epsilon$ expansion of the mode function with modified dispersion relation}

\begin{figure}[tb]
  \begin{center}
    \includegraphics[keepaspectratio=true,height=80mm]{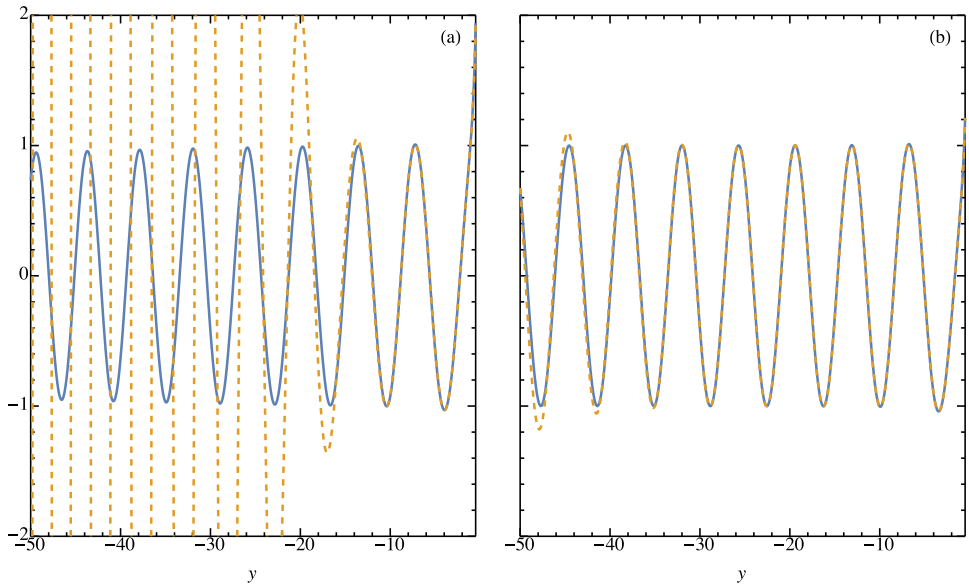}
  \end{center}
  \caption{The real part of the exact solution~(\ref{ex-Whitt}) (solid lines) versus
  that of the approximate expression~(\ref{approx}) (dashed lines) as functions of $y\,(=c_hk\eta)$.
  We take (a) $\delta=10^{-2}$ and (b) $\delta=10^{-3}$. It can be seen that
  the approximation is good for $\delta\times y^2\lesssim 1$.
  }%
  \label{fig:anvsap.eps}
\end{figure}

In the next section we compute the 3-point functions
by means of the in-in formalism~\cite{Maldacena:2002vr}.
To do so, one has to integrate products of $\psi_{\mathbf{k}}$ with respect to time.
Unfortunately, this turns out to be impossible in an analytic way if
one directly uses $\psi_{\mathbf{k}}$ written in terms of the Whittaker function.
Therefore, we assume that the modification to the dispersion relation is small,
and expand $\psi_{\mathbf{k}}$ in terms of $\epsilon$.
The approximation allows us to calculate the 3-point functions analytically
in a perturbative manner.

To second order in $\epsilon$, we find, up to a phase factor,
\begin{eqnarray}
	\frac{a{\sqrt{{\cal G}_T}}}{2}\psi_{\mathbf{k}}
	=F(\delta)\frac{e^{-iy+i\delta y^2/2}}{\sqrt{2c_h k}}
	\left[-\frac{i}{y}+1-\frac{\delta}{2}\left(y +iy^2\right)-\delta^2
	\left(\frac{5}{12}y^2+\frac{i}{24}y^3+\frac{1}{8}y^4 \right)+{\cal O}(\delta^3)\right]
	,\label{approx}
\end{eqnarray}
where $y:=c_hk\eta$ and $\delta:=\epsilon/c_h^2$.
The approximation is valid as long as $\delta\times y^2 = \epsilon k^2\eta^2\lesssim 1$,
as is clearly seen from Fig.~\ref{fig:anvsap.eps}.
We use this formula in the following calculations of the 3-point functions of the tensor perturbations.
The same approximation was used in Ref.~\cite{Ashoorioon:2011eg}
to evaluate the effects of nonstandard dispersion relations
on non-Gaussianities of the curvature perturbations.

\section{Cubic interactions and 3-point correlation functions}

\subsection{The cubic action}

Expanding Eq.~(\ref{eq:relevant action}) to third order in $h_{ij}$,
we obtain the following cubic action,
\begin{eqnarray}
S&=&\int\D t\D^3x\,a^3\biggl\{
\frac{c_3}{8}\dot h_i^j\dot h_j^k\dot h_k^i
+\frac{\mathcal{F}_T}{4a^2}
\left(h_{ik}h_{jl}-\frac{1}{2}h_{ij}h_{kl}\right)h_{ij,kl}
+\frac{a_7}{8a^4} \dot h_k^i \partial^2 h_j^k \partial^2 h_i^j
-\frac{b_6}{8a^2}\dot h_k^i \dot h_j^k \partial^2 h_i^j
\nonumber\\&&\qquad \qquad \quad
+\frac{\widetilde{d}_3}{a^4}\partial^2h_{ij}
\left[
\frac{1}{2}h_{ik,l}h_{jl,k}+h_{kl}\left(h_{ik,lj}-\frac{1}{4}h_{kl,ij}-\frac{1}{2}h_{ij,kl}\right)
\right]
-\frac{d_7}{8a^6}\partial^2 h_i^j\partial^2 h_j^k\partial^2 h_k^i
\biggr\}.\label{3action}
\end{eqnarray}

At this stage, several comments are in order.
It is easy to see that $\widetilde b_2\delta K_i^j\delta K_j^i$
does not contribute to the cubic action.
It is also noted that $\widetilde d_1 R$ and $\widetilde a_2 R_i^j\delta K_j^i$
give rise to the identical terms even at cubic order, leading to
the single combination having the coefficient ${\cal F}_T$.
We are thus left with the above 6 combinations out of the 8 terms in Eq.~(\ref{eq:relevant action}).
If the gravitational sector of the theory is described only by the Einstein-Hilbert term,
only the terms with the coefficient ${\cal F}_T$ remain, and therefore
all the other terms signal theories beyond general relativity.

Now let us look at the Horndeski and GLPV subclass by taking
$a_7=b_6=\widetilde d_3=d_7=0$:
\begin{eqnarray}
S\to\int\D t\D^3x\,a^3\biggl[
\frac{c_3}{8}\dot h_i^j\dot h_j^k\dot h_k^i
+\frac{\mathcal{F}_T}{4a^2}
\left(h_{ik}h_{jl}-\frac{1}{2}h_{ij}h_{kl}\right)h_{ij,kl}
\biggr].\label{H-GLPV}
\end{eqnarray}
As shown in Ref.~\cite{Gao:2011vs}, we have two independent cubic interactions
in the Horndeski theory. Those correspond to the two contributions displayed in the above action.
We see that no new contributions appear and
we have only the above two even in the GLPV theory beyond Horndeski.
Thus, we conclude that
{\em the tensor bispectrum in the GLPV theory reduces to a combination of
the shapes obtained within the Horndeski theory}~\cite{Gao:2011vs}.

The other four interaction terms newly appear in Gao's framework.
Among them the $\widetilde d_3$ and $d_7$ terms have also been studied
in the context of Ho\v{r}ava gravity~\cite{Huang:2013epa},
though the modification of the mode function due to the nonstandard dispersion relation
has been ignored for simplicity.
It is worth noting that in the case of the standard dispersion relation, $\widetilde d_3=0$,
the $b_6$ term can be recast in the form of the $c_3$ term
by the use of the linear equation of motion and integration by parts.

\subsection{3-point correlation functions}

We turn to evaluate the 3-point correlation functions of the tensor modes.
Although some of the 3-point functions have already been obtained in the literature~\cite{Gao:2011vs,Huang:2013epa},
below we present all the results for completeness. We also evaluate the corrections
arising from the nonstandard dispersion relation.
Using the in-in formalism~\cite{Maldacena:2002vr}, the 3-point functions can be
computed as
\begin{eqnarray}
	\langle
	\widetilde{h}_{i_1j_1}({\bf k}_1)\widetilde{h}_{i_2j_2}({\bf k}_2)\widetilde{h}_{i_3j_3}({\bf k}_3)
	\rangle
	=-i \int_{t_0}^t \D t' \left\langle\left[\widetilde{h}_{i_1j_1}(t,{\bf k}_1)
	\widetilde{h}_{i_2j_2}(t,{\bf k}_2)\widetilde{h}_{i_3j_3}(t,{\bf k}_3)
	,~H_\mathrm{int}(t') \right]\right\rangle,
\end{eqnarray}
where $H_{\rm int}$ is the interaction Hamiltonian obtained from the cubic action~(\ref{3action}).
Actually, integration with respect to time is to be performed using the conformal time rather than $t$
from $\eta=-\infty$ to $\eta=0$.
As mentioned earlier, this cannot be done analytically using the exact mode function
written in terms of the Whittaker function~(\ref{ex-Whitt}).
To make this step feasible, we use instead the approximate expression~(\ref{approx}).

For convenience we introduce the non-Gaussian amplitude ${\cal A}_{i_1j_1i_2j_2i_3j_3}$
defined by~\cite{Gao:2011vs}
\begin{eqnarray}
	\langle
	\widetilde{h}_{i_1j_1}({\bf k}_1)\widetilde{h}_{i_2j_2}({\bf k}_2)\widetilde{h}_{i_3j_3}({\bf k}_3)
	\rangle
	=(2\pi)^7 \delta^{(3)}(\mathbf{k}_1+\mathbf{k}_2+\mathbf{k}_3) \frac{\mathcal{P}_h^2}{k_1^3k_2^3k_3^3}
	\mathcal{A}_{i_1j_1i_2j_2i_3j_3},
\end{eqnarray}
where $\mathcal{A}_{i_1j_1i_2j_2i_3j_3}$ is written as a sum of each contribution,
which we compute to order $\epsilon^2$.
It turns out that the corrections due to the nonstandard dispersion relation
start at ${\cal O}(\epsilon^2)$, so that we write
\begin{align}
 \mathcal{A}_{i_1j_1i_2j_2i_3j_3}=\sum_{\bullet = c_3, a_7, \cdots} \left(
 \mathcal{A}^{(\bullet)}_{i_1j_1i_2j_2i_3j_3}+
 \frac{\epsilon^2}{c_h^4}\mathcal{C}^{(\bullet)}_{i_1j_1i_2j_2i_3j_3}
 \right),\label{A-divide}
\end{align}
where $\bullet = c_3, a_7, \cdots$ denotes the corresponding term in the cubic action.

First, the non-Gaussian amplitude arising from the cubic interaction
with the coefficient ${\cal F}_T$ is given by
\begin{align}
{\cal A}_{i_1j_1i_2j_2i_3j_3}^{({\cal F}_T)}&=
\widetilde{\cal A}
	\left\{
	\Pi_{i_1 j_1,ik}({\bf k}_1)\Pi_{i_2 j_2,jl}({\bf k}_2)
	\left[k_{3k}k_{3l}\Pi_{i_3 j_3,ij}({\bf k}_3)-\frac{1}{2}k_{3i}k_{3k}\Pi_{i_3 j_3,jl}({\bf k}_3)\right]
	+\mathrm{5~permutations~of~1,2,3}
	\right\},\label{eq:AGR}
	\\
{\cal C}_{i_1j_1i_2j_2i_3j_3}^{({\cal F}_T)}&=
\widetilde{\cal C}
	\biggl\{
	\Pi_{i_1 j_1,ik}({\bf k}_1)\cdots\biggr\},\label{eq:CGR}
\end{align}
where in ${\cal C}_{i_1j_1i_2j_2i_3j_3}^{({\cal F}_T)}$ the structure constructed from $\Pi_{ij,kl}$
is identical to that in ${\cal A}_{i_1j_1i_2j_2i_3j_3}^{({\cal F}_T)}$ and
\begin{align}
	\widetilde{\mathcal{A}}
	&=\frac{K_t}{16}
	\left(-1 +\frac{K_2}{K^2_t} +\frac{K_3}{K^3_t}\right),
	\\
\widetilde{\cal C}&=\frac{K_t}{8}
 \left(1-\frac{K_2}{K^2_t}+3\frac{K_3}{K^3_t} 
	- 2 \frac{K_2^2}{K^4_t} -6 \frac{K_2 K_3}{K^5_t}+6 \frac{K_3^2}{K^6_t}\right).
\end{align}
Here, we introduced the notations
$K_t:=k_1+k_2+k_3$, $K_2:=k_1 k_2+k_2 k_3+k_3 k_1$, and $K_3:= k_1 k_2 k_3$.
Note that ${\cal A}_{i_1j_1i_2j_2i_3j_3}^{({\cal F}_T)}$
is the only term present in the case of general relativity.
Interestingly, this term has the fixed and universal form
and is insensitive to an underlying theory and the inflationary energy scale.
This result strengthens
the statement originally made in the Horndeski theory in Ref.~\cite{Gao:2011vs}.

The concrete expressions for the other amplitudes are obtained as
\begin{align}
\mathcal{A}^{(c_3)}_{i_1j_1i_2j_2i_3j_3}
	&=
	\frac{3 c_3}{8}
	\frac{H}{\mathcal{G}_T}
	\cdot \frac{K_3^2}{K^3_t}
	\Pi_{123},\label{eq:Ac3}
	\\
\mathcal{A}^{(a_7)}_{i_1j_1i_2j_2i_3j_3}
	&=
	\frac{3 a_7}{8}
	\frac{H^3}{\mathcal{G}_T c_h^4}
	\cdot \frac{K_3^2}{K^3_t}
	\left(3+4\frac{K_2}{K^2_t}\right)\Pi_{123},
\label{eq:Aa7}
	\\
\mathcal{A}^{(b_6)}_{i_1j_1i_2j_2i_3j_3}
	&=
	-\frac{3 b_6}{4} 
	\frac{H^2}{\mathcal{G}_T c_h^2}
	\cdot \frac{K_3^2}{K^3_t}
	\Pi_{123},\label{eq:Ab6}
	\\
	\mathcal{A}^{(d_7)}_{i_1j_1i_2j_2i_3j_3}
	&=
	\frac{3 d_7}{2}
	\frac{H^4}{\mathcal{G}_T c_h^6}
	\cdot \frac{K_3^2}{K^3_t}
	\left(1+3\frac{K_2}{K^2_t}+15 \frac{K_3}{K^3_t}\right)\Pi_{123},\label{eq:Ad7}
\end{align}
and
\begin{align}
\mathcal{C}^{(c_3)}_{i_1j_1i_2j_2i_3j_3}
	&=
\mathcal{A}^{(c_3)}_{i_1j_1i_2j_2i_3j_3}\cdot 2 
\left(1-12 \frac{K_2}{K^2_t}+15 \frac{K_3}{K^3_t}\right)
,
\label{eq:Cc3}
	\\
\mathcal{C}^{(a_7)}_{i_1j_1i_2j_2i_3j_3}
	&=
 \mathcal{A}^{(a_7)}_{i_1j_1i_2j_2i_3j_3}\cdot 2
 \left(27-14 \frac{K_2}{K^2_t}+30\frac{K_3}{K^3_t} 
	- 240 \frac{K_2^2}{K^4_t} +210 \frac{K_2 K_3}{K^5_t}\right)
	\left(3+4\frac{K_2}{K_t^2}\right)^{-1}
	,
\label{eq:Ca7}
	\\
\mathcal{C}^{(b_6)}_{i_1j_1i_2j_2i_3j_3}
	&=
\mathcal{A}^{(b_6)}_{i_1j_1i_2j_2i_3j_3}\cdot 2
\left(7-29 \frac{K_2}{K^2_t}+30\frac{K_3}{K^3_t}\right)
	,\label{eq:Cb6}
	\\
	\mathcal{C}^{(d_7)}_{i_1j_1i_2j_2i_3j_3}
	&=
\mathcal{A}^{(d_7)}_{i_1j_1i_2j_2i_3j_3}\cdot 20
\left(1+3 \frac{K_2}{K^2_t}+51\frac{K_3}{K^3_t} 
	- 18 \frac{K_2^2}{K^4_t} -126 \frac{K_2 K_3}{K^5_t}+126 \frac{K_3^2}{K^6_t}\right)
	\left(1+3\frac{K_2}{K^2_t}+15 \frac{K_3}{K^3_t}\right)^{-1}
	,\label{eq:Cd7}
\end{align}
where we introduced the shortened notation for the common factor:
$\Pi_{123}:=\Pi_{i_1j_1,ij}({\bf k}_1) \Pi_{i_2j_2,jk}({\bf k}_2) \Pi_{i_3j_3,ki}({\bf k}_3)$.
We see that ${\cal A}_{i_1j_1i_2j_2i_3j_3}^{(c_3)}$ and ${\cal A}_{i_1j_1i_2j_2i_3j_3}^{(b_6)}$
have the same momentum dependences as expected.
Finally,
the $\widetilde d_3$ term itself is a small correction of ${\cal O}(\epsilon^2)$ in our approximation,
so that ${\cal A}_{i_1j_1i_2j_2i_3j_3}^{(\widetilde d_3)} =0$ and
\begin{align}
\mathcal{C}^{(\widetilde d_3)}_{i_1j_1i_2j_2i_3j_3}&=
	\frac{1}{4K_t}\left[1 +\frac{K_2}{K^2_t} +3\frac{K_3}{K^3_t}\right]
	\biggl\{-k_3^2\Pi_{i_3j_3,ij}(\mathbf{k}_3)
	\biggl[\frac{1}{2}\Pi_{i_2j_2,jl}(\mathbf{k}_2) \Pi_{i_1j_1,ik}(\mathbf{k}_1)k_{2l}k_{1k}
	\notag\\& \quad
	+\Pi_{i_2j_2,kl}
	\left(\Pi_{i_1j_1,ik}k_{1l}k_{1j}-\frac{1}{4}\Pi_{i_1j_1,kl}k_{1i}k_{1j}
	-\frac{1}{2}\Pi_{i_1j_1,ij}k_{1k}k_{1l}\right)\biggr]
	+\mathrm{5~permutations~of~1,2,3}\bigg\}.\label{eq:Ad3}
\end{align}

Let us investigate the 3-point correlation functions
between each polarization mode of gravitational waves.
The two polarization modes are expressed as
\begin{eqnarray}
	\xi^{(s)}({\bf k})=\tilde h_{ij}({\bf k}) e_{ij}^{\ast (s)}({\bf k}).
\end{eqnarray}
For the 3-point functions $\langle \xi^{(s_1)}\xi^{(s_2)}\xi^{(s_3)}\rangle$
its amplitude can be evaluated by computing
\begin{eqnarray}
\left\{
\begin{array}{c}
\mathcal A^{s_1s_2s_3}_{(\bullet)} \\
\mathcal C^{s_1s_2s_3}_{(\bullet)} 
\end{array} \right\}
=
	e_{i_1j_1}^{\ast (s_1)}({\bf k}_1)
	e_{i_2j_2}^{\ast (s_2)}({\bf k}_2)
	e_{i_3j_3}^{\ast (s_3)}({\bf k}_3)\times
	\left\{
\begin{array}{c}
\mathcal A_{i_1j_1 i_2j_2 i_3j_3}^{(\bullet)}\\
\mathcal C_{i_1j_1 i_2j_2 i_3j_3}^{(\bullet)}
\end{array} \right\}.
\end{eqnarray}
It is straightforward to obtain the following amplitudes,
\begin{align}
	\mathcal{A}^{s_1 s_2 s_3}_{({\cal F}_T)}
	&=
	\widetilde{\mathcal{A}}
	\cdot F^{s_1 s_2 s_3}_{({\rm GR})},
	\\
	\mathcal{A}^{s_1 s_2 s_3}_{(c_3)}
	&=
	\frac{3 c_3}{8} 
	\frac{H}{\mathcal{G}_T }
	\cdot \frac{K_3^2}{K^3_t}
	F^{s_1 s_2 s_3}_{({\rm H})},
	\\
	\mathcal{A}^{s_1 s_2 s_3}_{(a_7)}
	&=
	\frac{3 a_7}{8}
	\frac{H^3}{\mathcal{G}_T c_h^4}
	\cdot \frac{K_3^2}{K^3_t}
	\left(3+4\frac{K_2}{K^2_t}\right)
	F^{s_1 s_2 s_3}_{({\rm H})},
	\\
	\mathcal{A}^{s_1 s_2 s_3}_{(b_6)}
	&=
	-\frac{3 b_6}{4} 
	\frac{H^2}{\mathcal{G}_T c_h^2}
	\cdot \frac{K_3^2}{K^3_t}
	F^{s_1 s_2 s_3}_{({\rm H})},
	\\
	\mathcal{A}^{s_1 s_2 s_3}_{(d_7)}
	&=
	\frac{3 d_7}{2}
	\frac{H^4}{\mathcal{G}_T c_h^6}
	\cdot \frac{K_3^2}{K^3_t}
	\left(1+3\frac{K_2}{K^2_t}+15 \frac{K_3}{K^3_t}\right)
	F^{s_1 s_2 s_3}_{({\rm H})},
\end{align}
and
\begin{align}
\mathcal{C}^{s_1 s_2 s_3}_{{\cal F}_T}
&=
\widetilde {\cal C}\cdot F_{({\rm GR})}^{s_1s_2s_3},
\\
	\mathcal{C}^{s_1 s_2 s_3}_{(c_3)}
	&=
\mathcal{A}^{s_1 s_2 s_3}_{(c_3)}\cdot 2
\left(1-12 \frac{K_2}{K^2_t}+15 \frac{K_3}{K^3_t}\right)
,
	\\
	\mathcal{C}^{s_1 s_2 s_3}_{(a_7)}
	&=
 \mathcal{A}^{s_1 s_2 s_3}_{(a_7)}\cdot 2
 \left(27-14 \frac{K_2}{K^2_t}+30\frac{K_3}{K^3_t} 
	- 240 \frac{K_2^2}{K^4_t} +210 \frac{K_2 K_3}{K^5_t}\right)
	\left(3+4\frac{K_2}{K_t^2}\right)^{-1}
,
	\\
	\mathcal{C}^{s_1 s_2 s_3}_{(b_6)}
	&=
\mathcal{A}^{s_1 s_2 s_3}_{(b_6)}\cdot 2
	\left(7-29 \frac{K_2}{K^2_t}+30\frac{K_3}{K^3_t}\right)
,
	\\
	\mathcal{C}^{s_1 s_2 s_3}_{(d_7)}
	&=
	\mathcal{A}^{s_1 s_2 s_3}_{(d_7)}\cdot 20
 \left(1+3 \frac{K_2}{K^2_t}+51\frac{K_3}{K^3_t} 
	- 18 \frac{K_2^2}{K^4_t} -126 \frac{K_2 K_3}{K^5_t}+126 \frac{K_3^2}{K^6_t}\right)
		\left(1+3\frac{K_2}{K^2_t}+15 \frac{K_3}{K^3_t}\right)^{-1}
,
	\\
	\mathcal{C}^{s_1 s_2 s_3}_{(\widetilde d_3)}
	&=
	\frac{1}{4K_t}\left(1 +\frac{K_2}{K^2_t} +3\frac{K_3}{K^3_t}\right)
	F^{s_1 s_2 s_3}_{(\widetilde d_3)},
\end{align}
where we defined
\begin{eqnarray}
	F^{+++}_{({\rm GR})}
	&:=&
	\frac{1}{128} \cdot \frac{K^8_t}{K_3^2}\left(1 -4\frac{K_2}{K^2_t} +8\frac{K_3}{K^3_t}\right),
	\\
		F^{+++}_{({\rm H})}
	&:=&
	\frac{1}{64} \cdot \frac{K^6_t}{K_3^2}\left(1 -4\frac{K_2}{K^2_t} +8\frac{K_3}{K^3_t}\right),
	\\
	F^{+++}_{({\widetilde d_3})}
	&:=&
	\frac{1}{256} \cdot \frac{K^{10}_t}{K_3^2}
	\left(1 -10\frac{K_2}{K^2_t} +28\frac{K_3}{K^3_t} +24\frac{K_2^2}{K^4_t}
	-128\frac{K_2 K_3}{K^5_t} +160\frac{K_3^2}{K^6_t}\right),
\end{eqnarray}
and when spins are mixed the corresponding functions can be derived from
$F^{++-}_{({\rm GR}, {\rm H}, \widetilde d_3)}(k_1,k_2,k_3)=F^{+++}_{({\rm GR}, {\rm H}, \widetilde d_3)}(k_1,k_2,-k_3)$.
Note the relations
\begin{eqnarray}
	F^{+++}_{({\rm GR})}&=&\frac{K^2_t}{2}F^{+++}_{({\rm H})}, 
	\\
	F^{+++}_{({\widetilde d_3})}
	&=&\frac{K^{10}_t}{4}\left(1-6\frac{K_2}{K^2_t}+20\frac{K_3}{K^3_t}\right)F^{+++}_{({\rm H})}.
\end{eqnarray}
Since theories in Gao's framework do not violate the parity symmetry, we have the properties
$F^{++-}_{({\rm GR}, {\rm H}, \widetilde d_3)}=F^{--+}_{({\rm GR}, {\rm H}, \widetilde d_3)}$ and
$F^{---}_{({\rm GR}, {\rm H}, \widetilde d_3)}=F^{+++}_{({\rm GR}, {\rm H}, \widetilde d_3)}$.
In the equilateral configuration, we have 
\begin{align}
	&F^{++-}_{({\rm H})}=F^{+-+}_{({\rm H})}=F^{-++}_{({\rm H})}=\frac{1}{9}F^{+++}_{({\rm H})},
	\\
	&F^{++-}_{({\rm GR})}=F^{+-+}_{({\rm GR})}=F^{-++}_{({\rm GR})}=\frac{1}{81}F^{+++}_{({\rm GR})},
	\\
	&F^{++-}_{(\widetilde d_3)}=F^{+-+}_{(\widetilde d_3)}=F^{-++}_{(\widetilde d_3)}
	=\frac{13}{189}F^{+++}_{(\widetilde d_3)},
\end{align}
while in the squeezed limit ${\bf k}_3\to 0$ we find
\begin{align}
	&F_{({\rm GR},\widetilde d_3)}^{+++} \approx  F_{({\rm GR},\widetilde d_3)}^{++-}\approx -\frac{k_1^2}{2},
	\\
	&F_{({\rm GR})}^{+-+} \approx F_{({\rm GR})}^{-++}\approx -\frac{k_3^4}{32k_1^2},
	\\
	&F_{(\widetilde d_3)}^{+-+} \approx  F_{(\widetilde d_3)}^{-++}\approx -\frac{7k_3^4}{32k_1^2}.
\end{align}

\begin{figure}[H]
\centering\includegraphics*[width=6.0cm,keepaspectratio,clip]{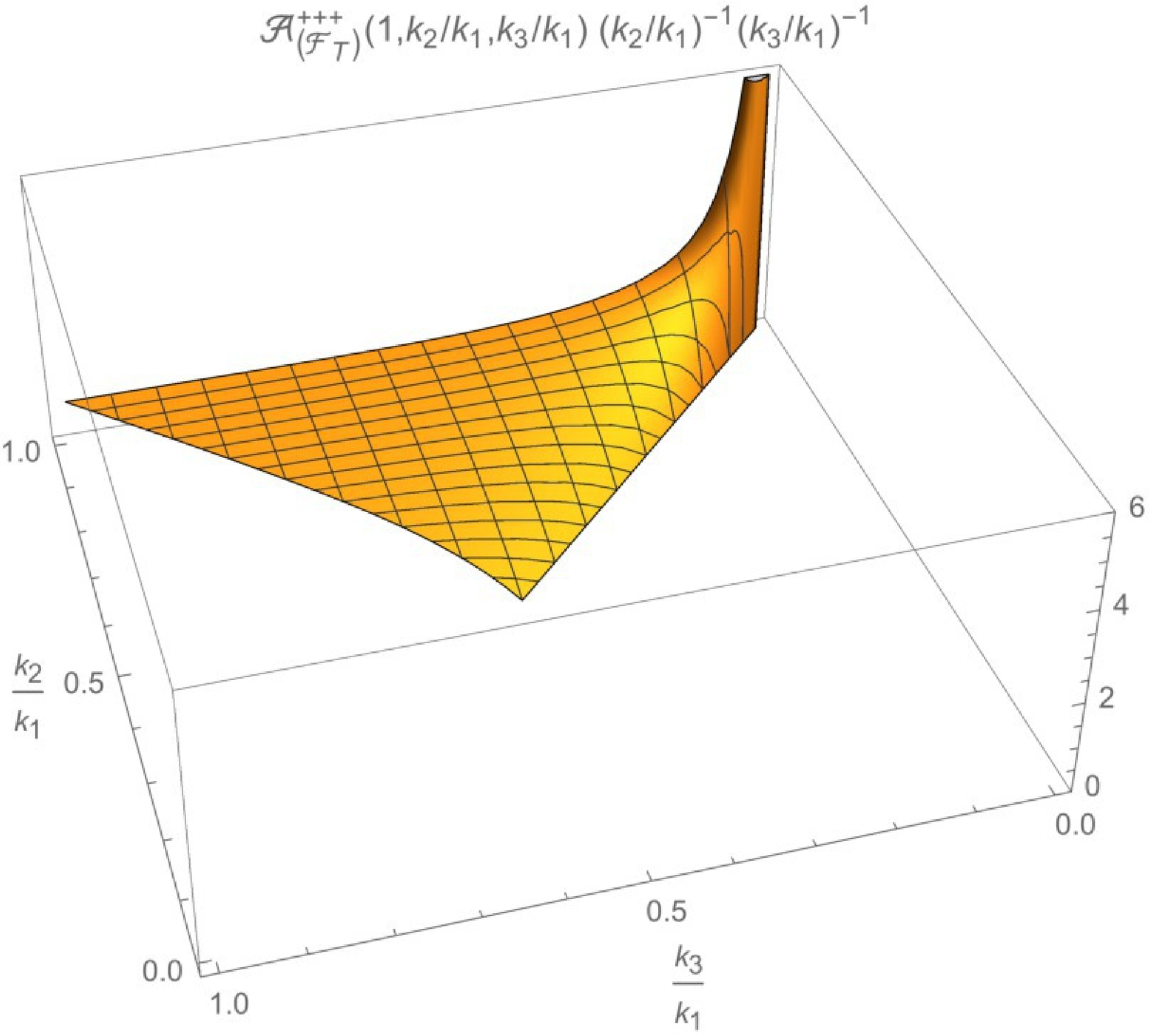}
\caption{$\mathcal A^{+++}_{({\cal F}_T)}(1,k_2/k_1,k_3/k_1) (k_2/k_1)^{-1}(k_3/k_1)^{-1}$
as a function of $k_2/k_1$ and $k_3/k_1$. The plot is normalized to 1
for equilateral configurations $k_2/k_1=k_3/k_1=1$.	}
\label{fig:AGR+++}
\end{figure}
\begin{figure}[H]
\centering\includegraphics*[width=6.0cm,keepaspectratio,clip]{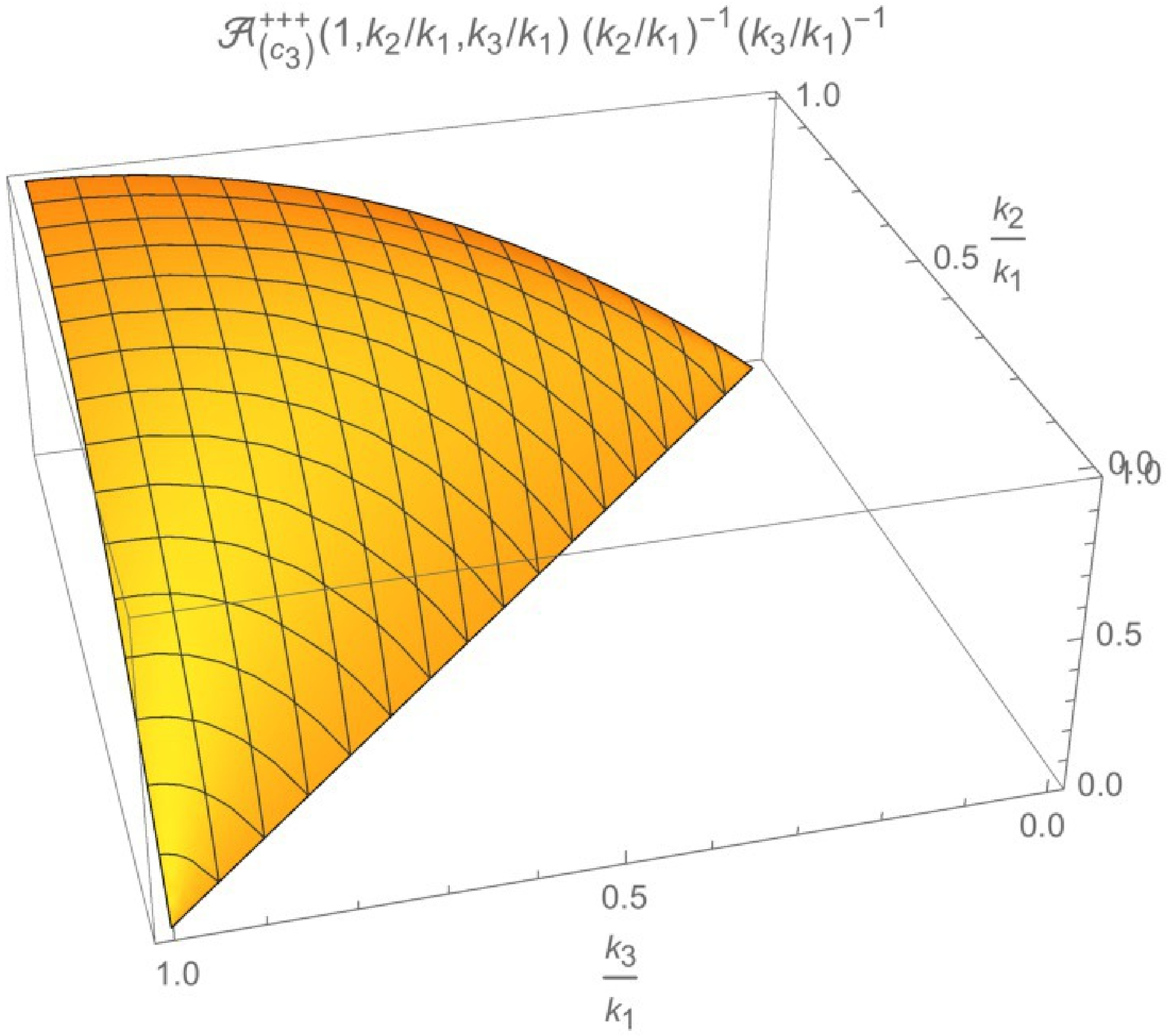}
\caption{$\mathcal A^{+++}_{(c_3)}(1,k_2/k_1,k_3/k_1) (k_2/k_1)^{-1}(k_3/k_1)^{-1}$
as a function of $k_2/k_1$ and $k_3/k_1$. The plot is normalized to 1
for equilateral configurations $k_2/k_1=k_3/k_1=1$.	}
\label{fig:Ac3+++}
\end{figure}

\begin{figure}[H]
\centering\includegraphics*[width=6.0cm,keepaspectratio,clip]{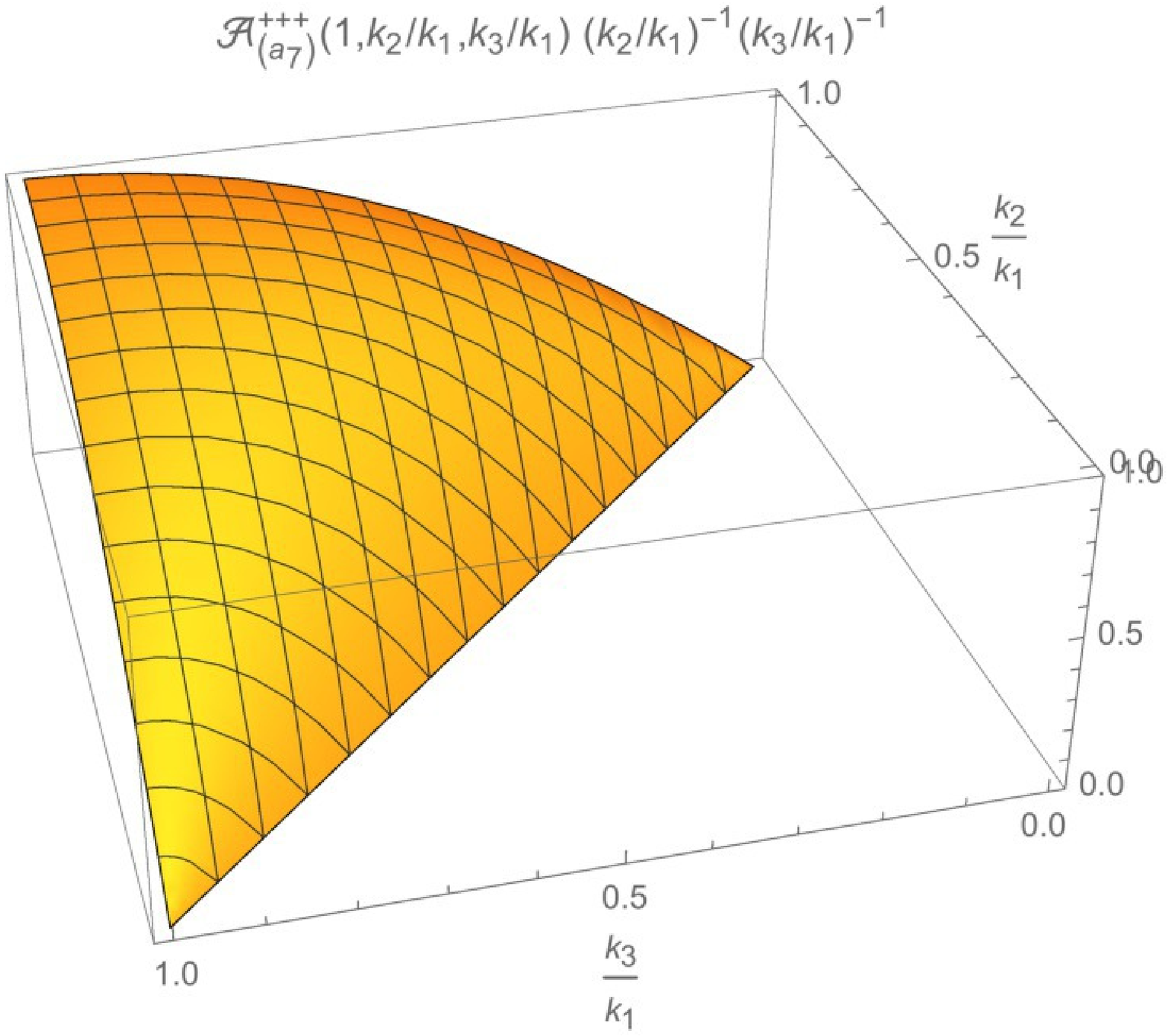}
\caption{$\mathcal A^{+++}_{(a_7)}(1,k_2/k_1,k_3/k_1) (k_2/k_1)^{-1}(k_3/k_1)^{-1}$
as a function of $k_2/k_1$ and $k_3/k_1$. The plot is normalized to 1
for equilateral configurations $k_2/k_1=k_3/k_1=1$.	}
\label{fig:Aa7+++}
\end{figure}
\begin{figure}[H]
\centering\includegraphics*[width=6.0cm,keepaspectratio,clip]{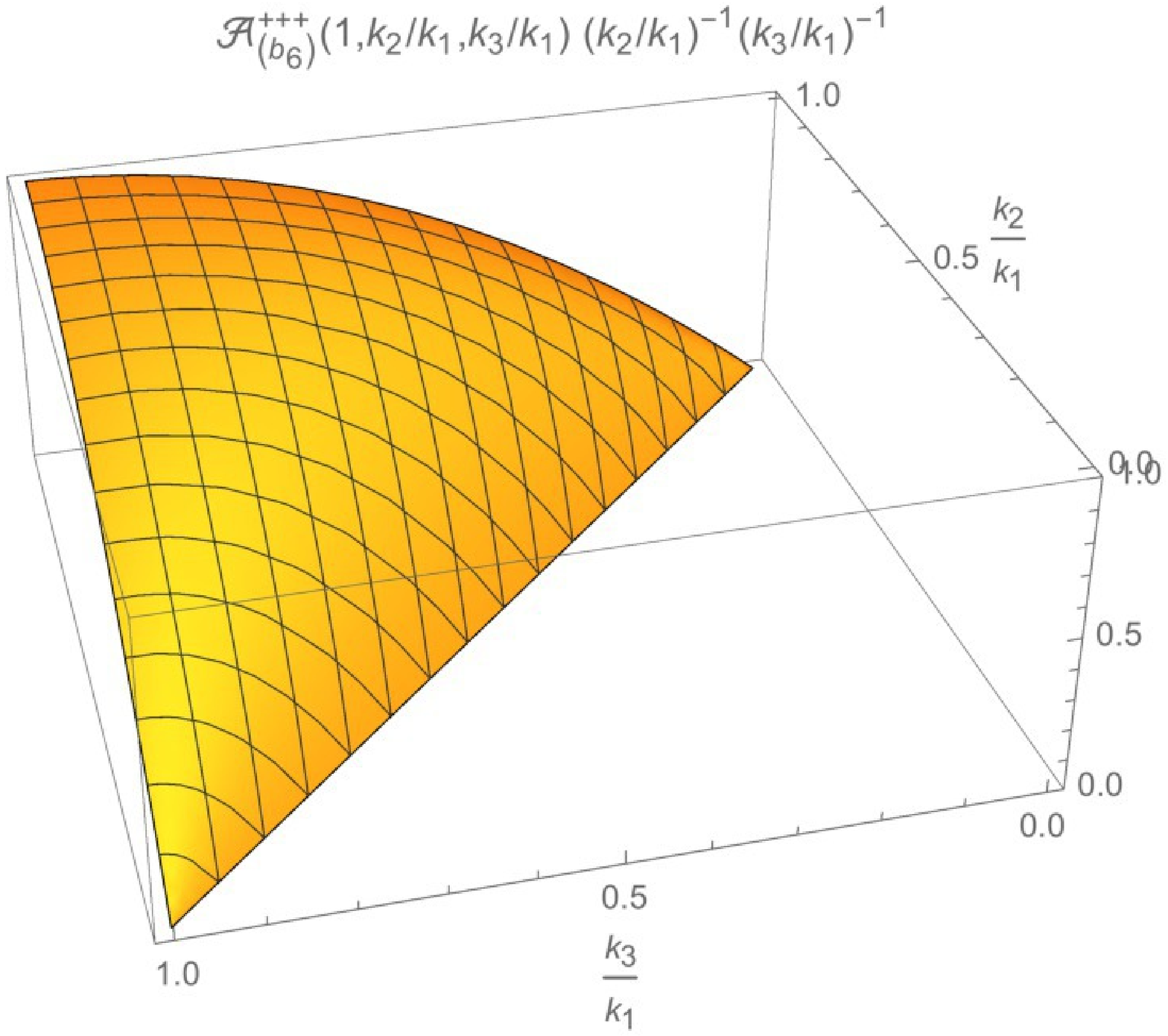}
\caption{$\mathcal A^{+++}_{(b_6)}(1,k_2/k_1,k_3/k_1) (k_2/k_1)^{-1}(k_3/k_1)^{-1}$
as a function of $k_2/k_1$ and $k_3/k_1$. The plot is normalized to 1
for equilateral configurations $k_2/k_1=k_3/k_1=1$.}
\label{fig:Ab6+++}
\end{figure}

\begin{figure}[H]
\centering\includegraphics*[width=6.0cm,keepaspectratio,clip]{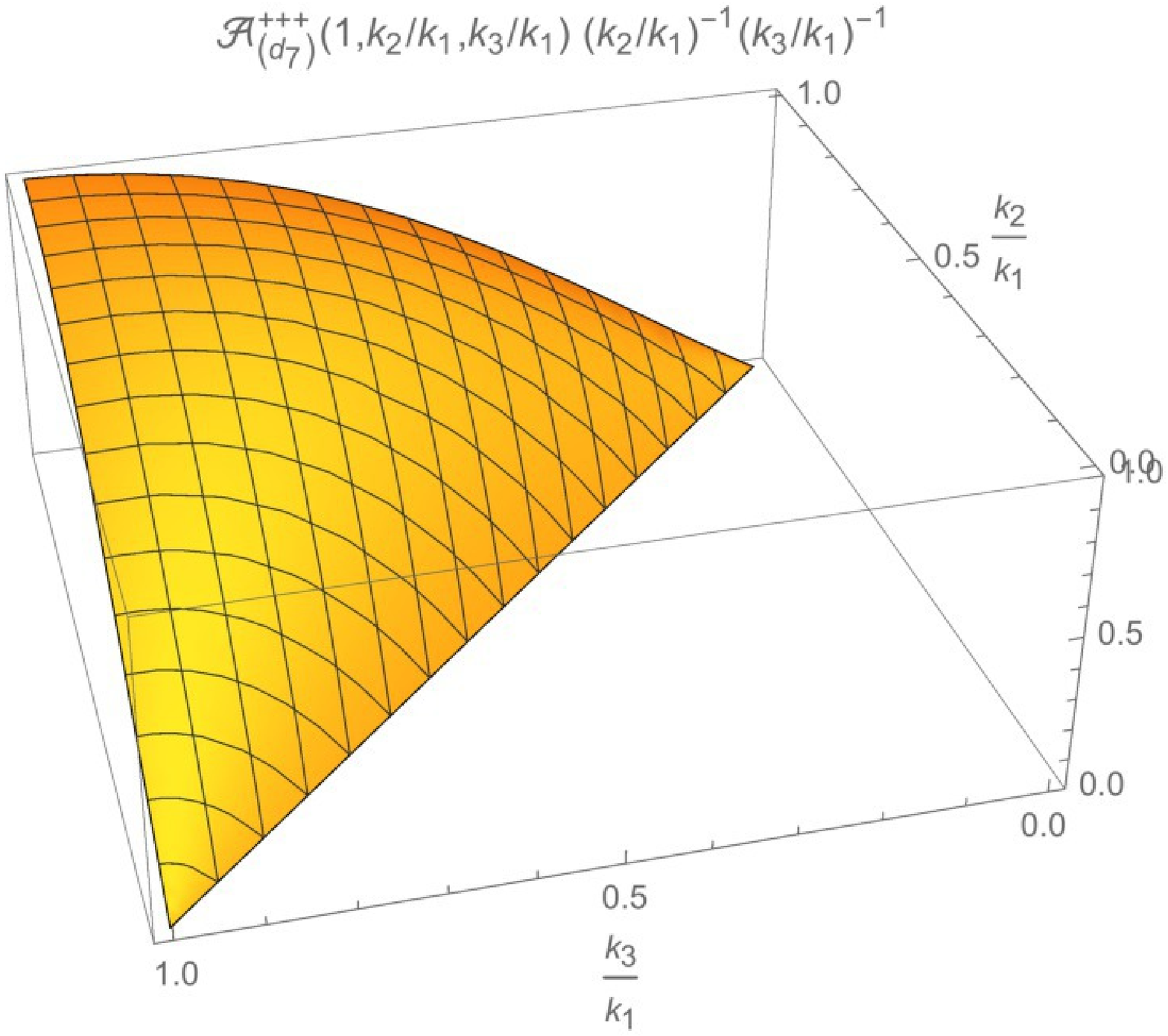}
\caption{$\mathcal A^{+++}_{(d_7)}(1,k_2/k_1,k_3/k_1) (k_2/k_1)^{-1}(k_3/k_1)^{-1}$
as a function of $k_2/k_1$ and $k_3/k_1$. The plot is normalized to 1
for equilateral configurations $k_2/k_1=k_3/k_1=1$.	}
\label{fig:Ad7+++}
\end{figure}
\begin{figure}[H]
\centering\includegraphics*[width=6.0cm,keepaspectratio,clip]{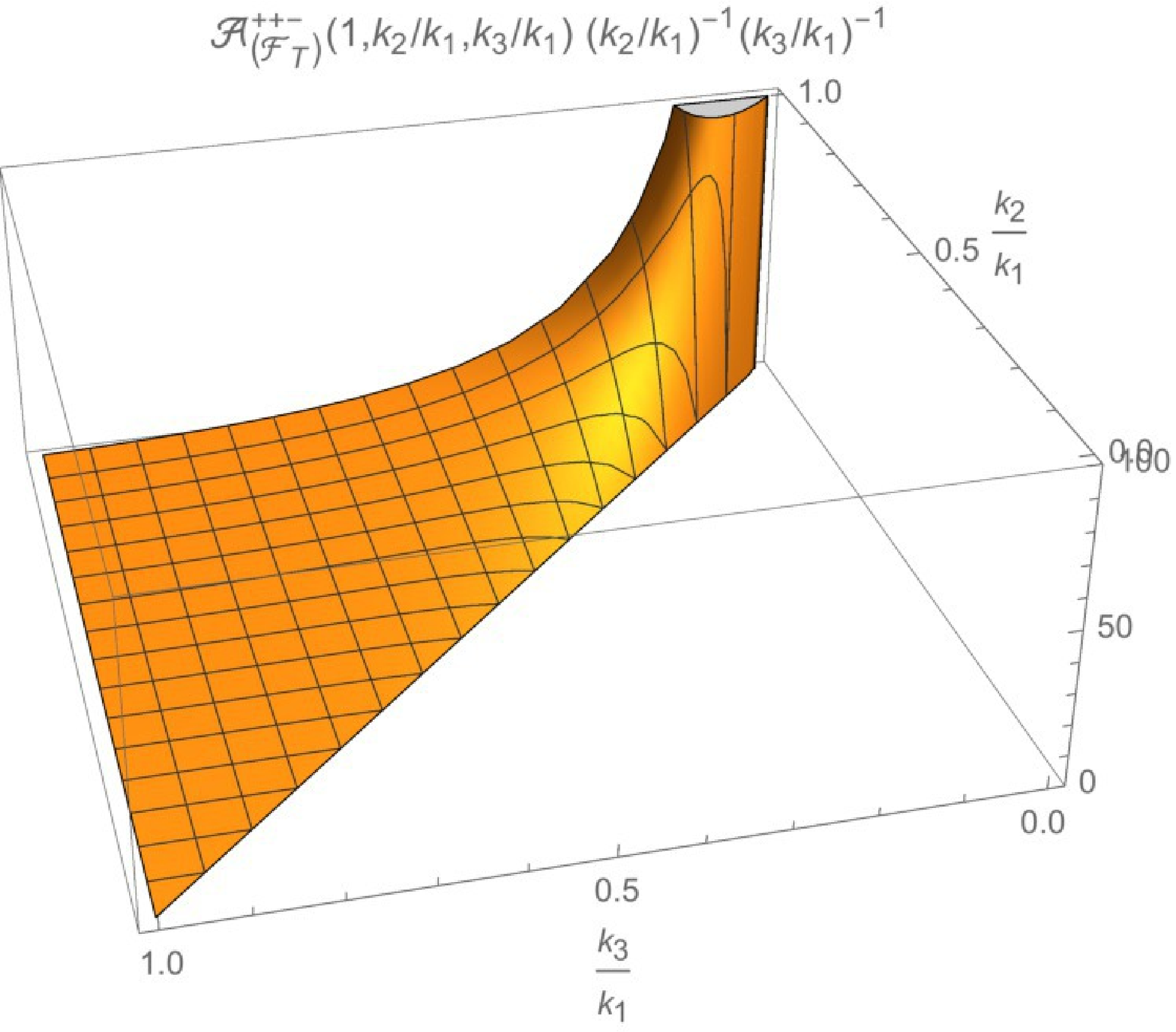}
\caption{$\mathcal A^{++-}_{({\cal F}_T)}(1,k_2/k_1,k_3/k_1) (k_2/k_1)^{-1}(k_3/k_1)^{-1}$
as a function of $k_2/k_1$ and $k_3/k_1$. The plot is normalized to 1
for equilateral configurations $k_2/k_1=k_3/k_1=1$.	}
\label{fig:AGR++-}
\end{figure}
\begin{figure}[H]
\centering\includegraphics*[width=6.0cm,keepaspectratio,clip]{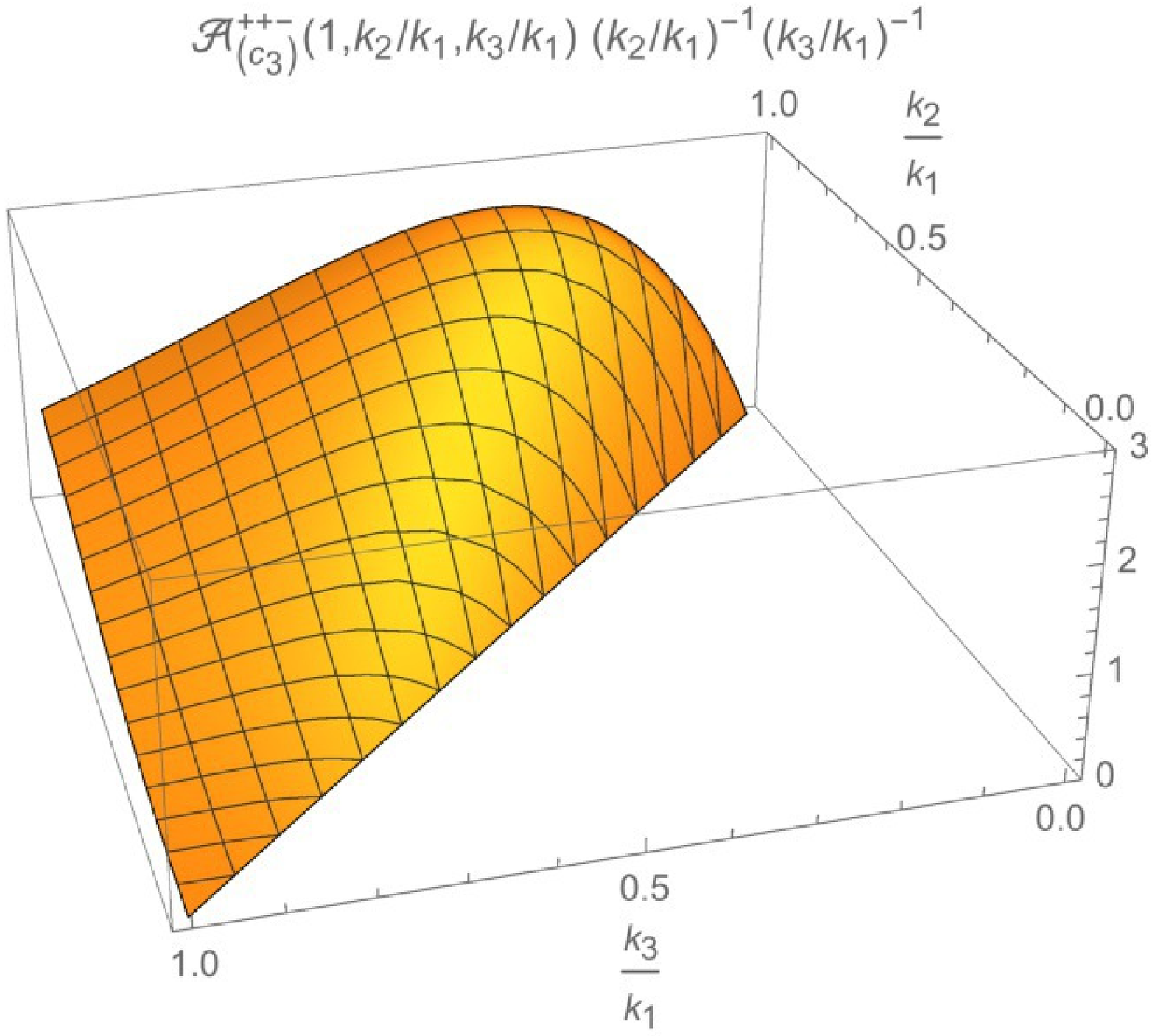}
\caption{$\mathcal A^{++-}_{(c_3)}(1,k_2/k_1,k_3/k_1) (k_2/k_1)^{-1}(k_3/k_1)^{-1}$
as a function of $k_2/k_1$ and $k_3/k_1$. The plot is normalized to 1
for equilateral configurations $k_2/k_1=k_3/k_1=1$.	}
\label{fig:Ac3++-}
\end{figure}

\begin{figure}[H]
\centering\includegraphics*[width=6.0cm,keepaspectratio,clip]{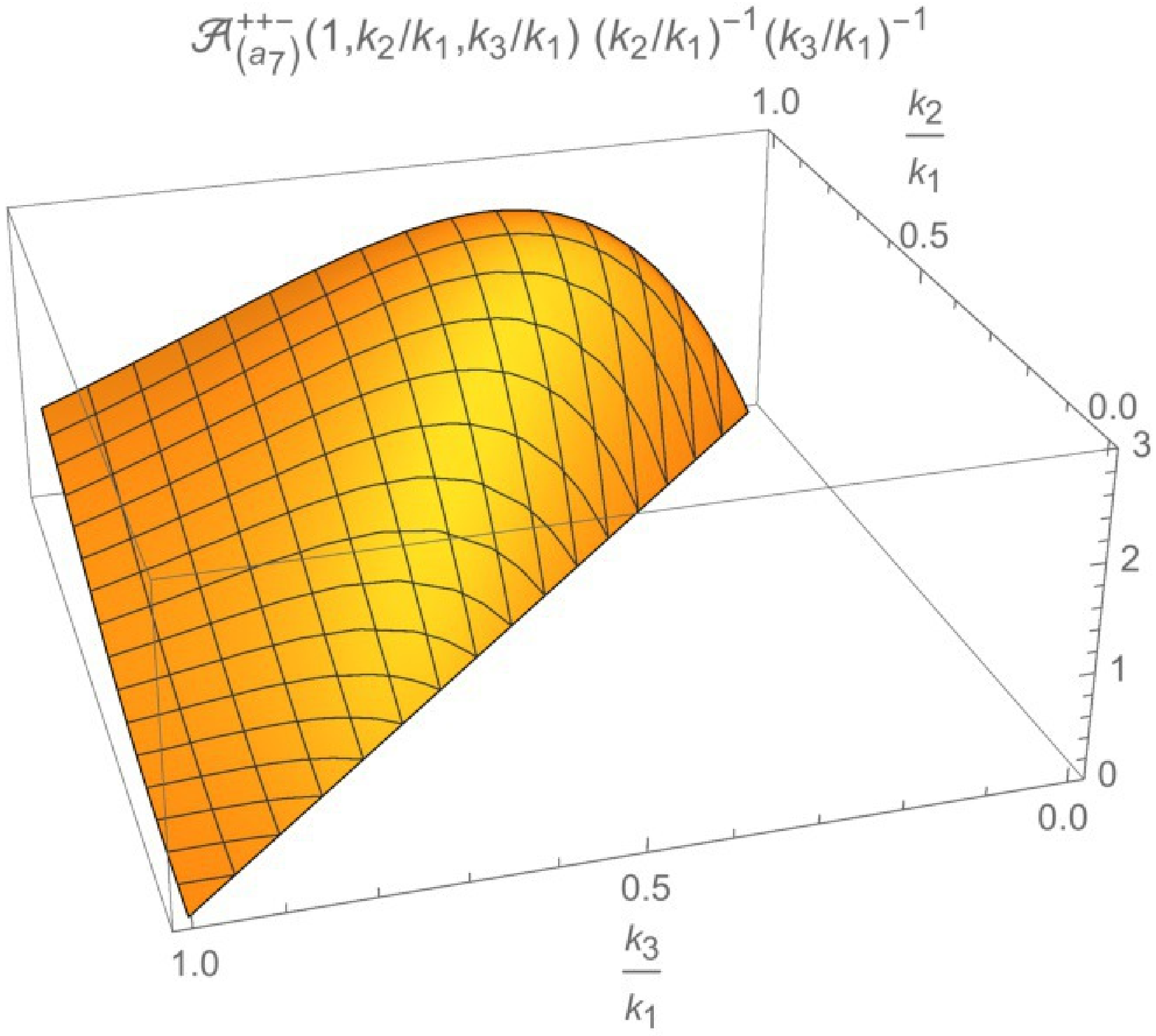}
\caption{$\mathcal A^{++-}_{(a_7)}(1,k_2/k_1,k_3/k_1) (k_2/k_1)^{-1}(k_3/k_1)^{-1}$
as a function of $k_2/k_1$ and $k_3/k_1$. The plot is normalized to 1
for equilateral configurations $k_2/k_1=k_3/k_1=1$.	}
\label{fig:Aa7++-}
\end{figure}
\begin{figure}[H]
\centering\includegraphics*[width=6.0cm,keepaspectratio,clip]{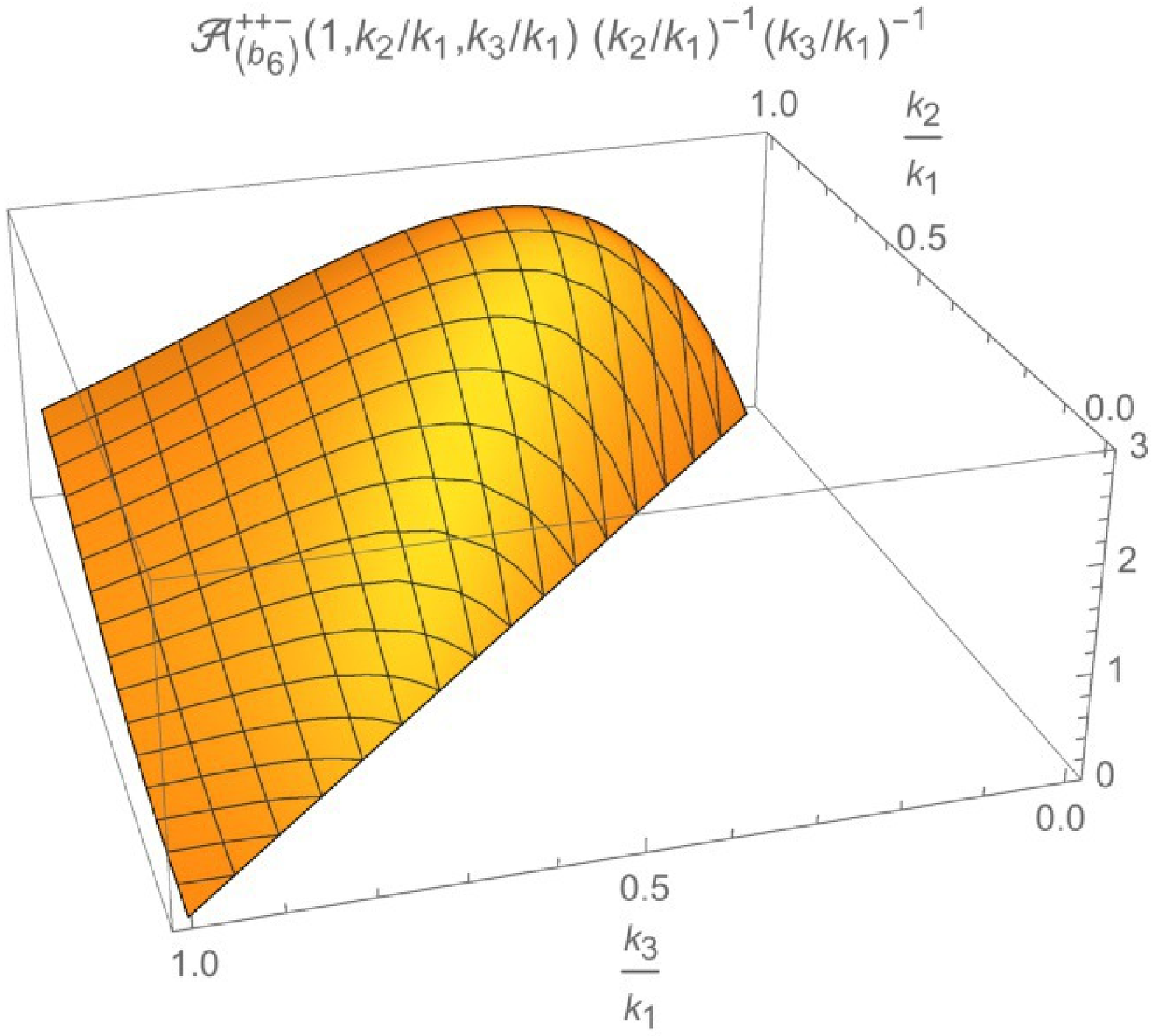}
\caption{$\mathcal A^{++-}_{(b_6)}(1,k_2/k_1,k_3/k_1) (k_2/k_1)^{-1}(k_3/k_1)^{-1}$
as a function of $k_2/k_1$ and $k_3/k_1$. The plot is normalized to 1
for equilateral configurations $k_2/k_1=k_3/k_1=1$.	}
\label{fig:Ab6++-}
\end{figure}

\begin{figure}[H]
\centering\includegraphics*[width=6.0cm,keepaspectratio,clip]{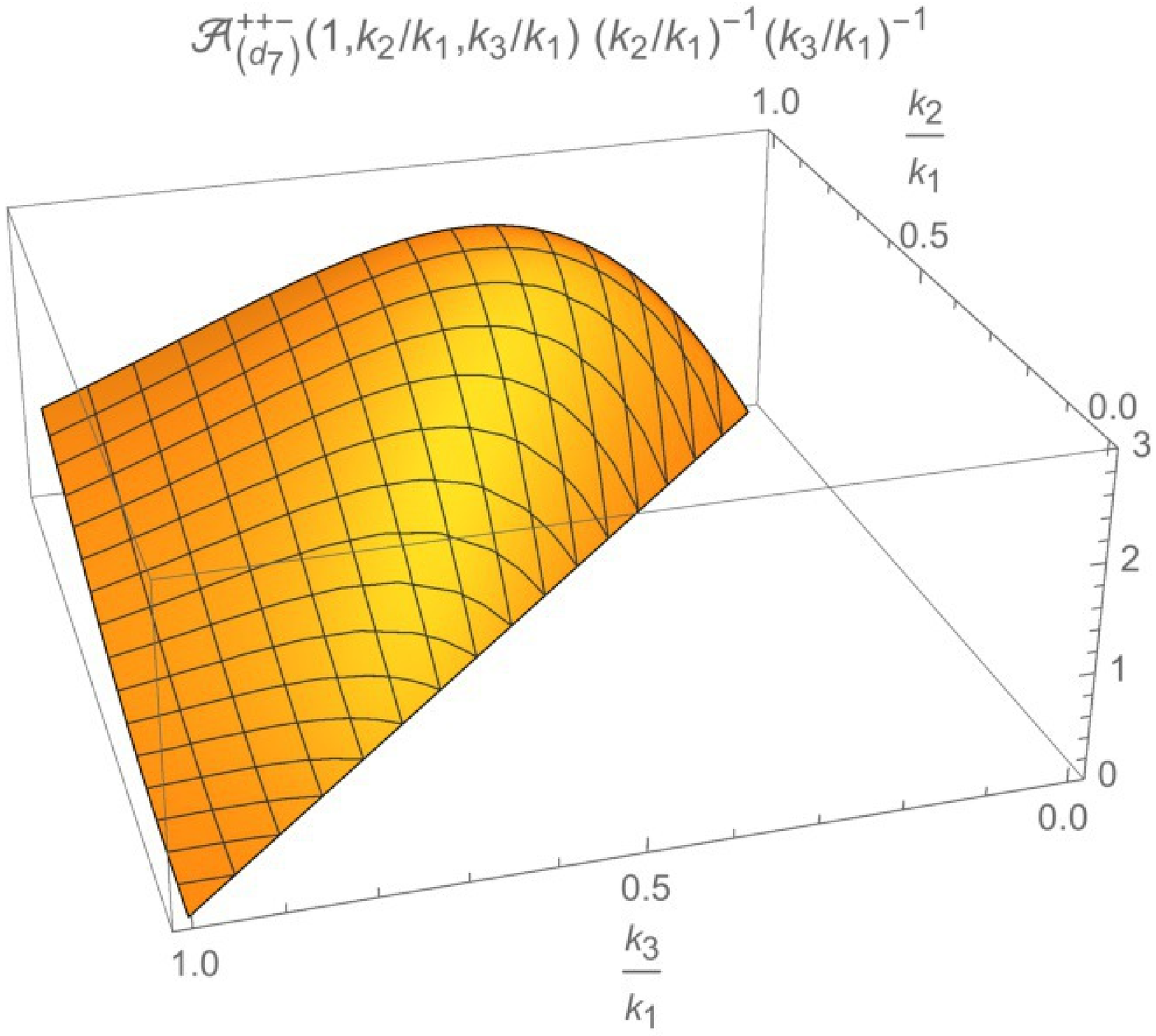}
\caption{$\mathcal A^{++-}_{(d_7)}(1,k_2/k_1,k_3/k_1) (k_2/k_1)^{-1}(k_3/k_1)^{-1}$
as a function of $k_2/k_1$ and $k_3/k_1$. The plot is normalized to 1
for equilateral configurations $k_2/k_1=k_3/k_1=1$.	}
\label{fig:Ad7++-}
\end{figure}

\begin{figure}[H]
\centering\includegraphics*[width=6.0cm,keepaspectratio,clip]{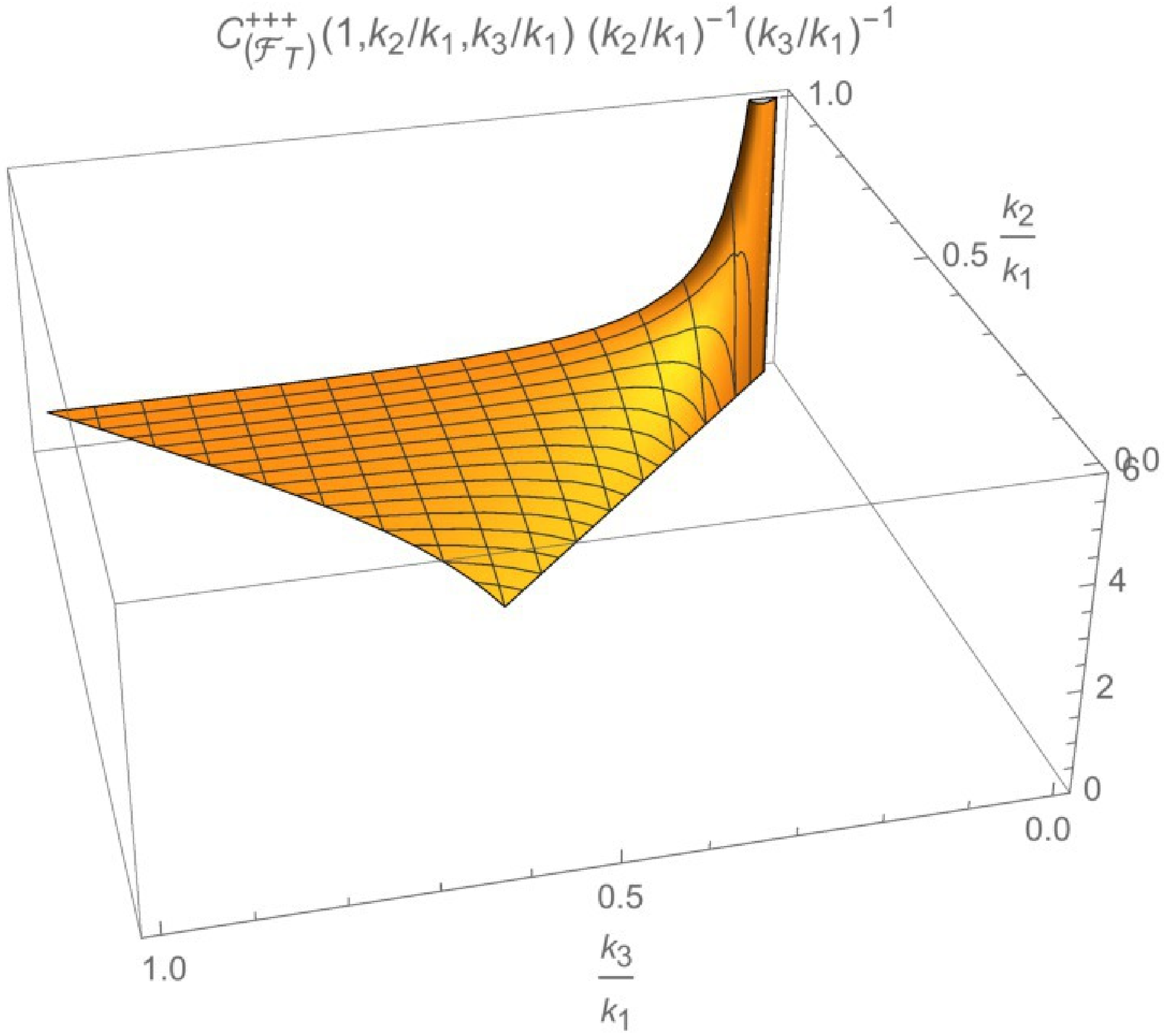}
\caption{$\mathcal C^{+++}_{({\cal F}_T)}(1,k_2/k_1,k_3/k_1) (k_2/k_1)^{-1}(k_3/k_1)^{-1}$
as a function of $k_2/k_1$ and $k_3/k_1$. The plot is normalized to 1
for equilateral configurations $k_2/k_1=k_3/k_1=1$.	}
\label{fig:CGR+++}
\end{figure}
\begin{figure}[H]
\centering\includegraphics*[width=6.0cm,keepaspectratio,clip]{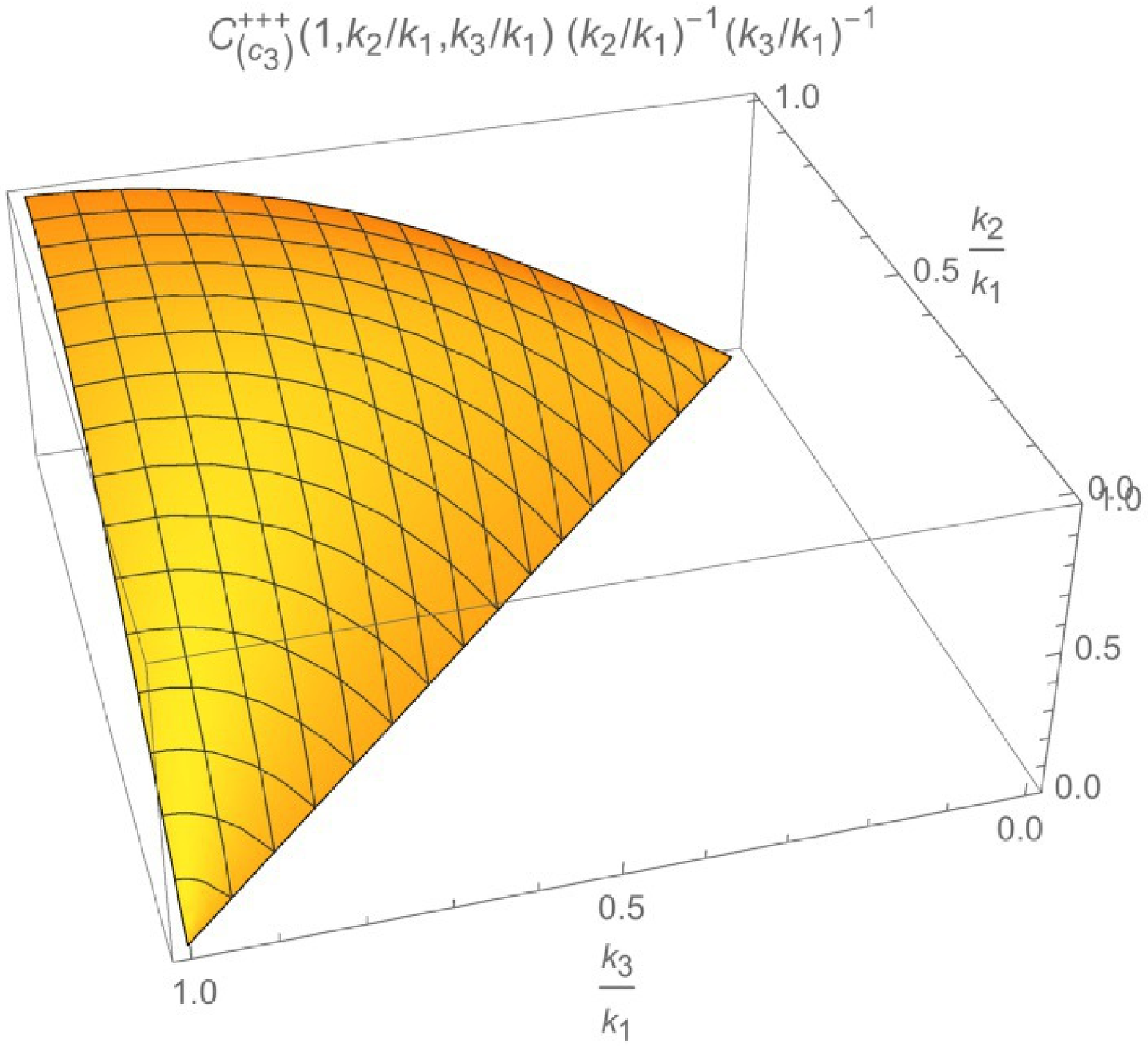}
\caption{$\mathcal C^{+++}_{(c_3)}(1,k_2/k_1,k_3/k_1) (k_2/k_1)^{-1}(k_3/k_1)^{-1}$
as a function of $k_2/k_1$ and $k_3/k_1$. The plot is normalized to 1
for equilateral configurations $k_2/k_1=k_3/k_1=1$.	}
\label{fig:Cc3+++}
\end{figure}

\begin{figure}[H]
\centering\includegraphics*[width=6.0cm,keepaspectratio,clip]{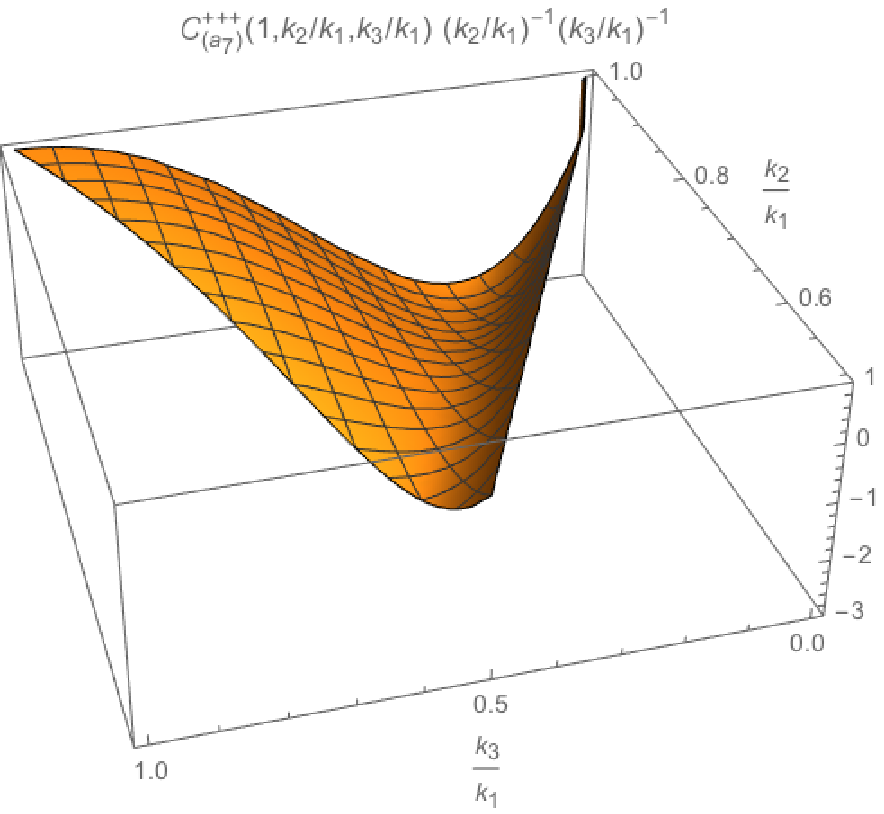}
\caption{$\mathcal C^{+++}_{(a_7)}(1,k_2/k_1,k_3/k_1) (k_2/k_1)^{-1}(k_3/k_1)^{-1}$
as a function of $k_2/k_1$ and $k_3/k_1$. The plot is normalized to 1
for equilateral configurations $k_2/k_1=k_3/k_1=1$.	}
\label{fig:Ca7+++}
\end{figure}
\begin{figure}[H]
\centering\includegraphics*[width=6.0cm,keepaspectratio,clip]{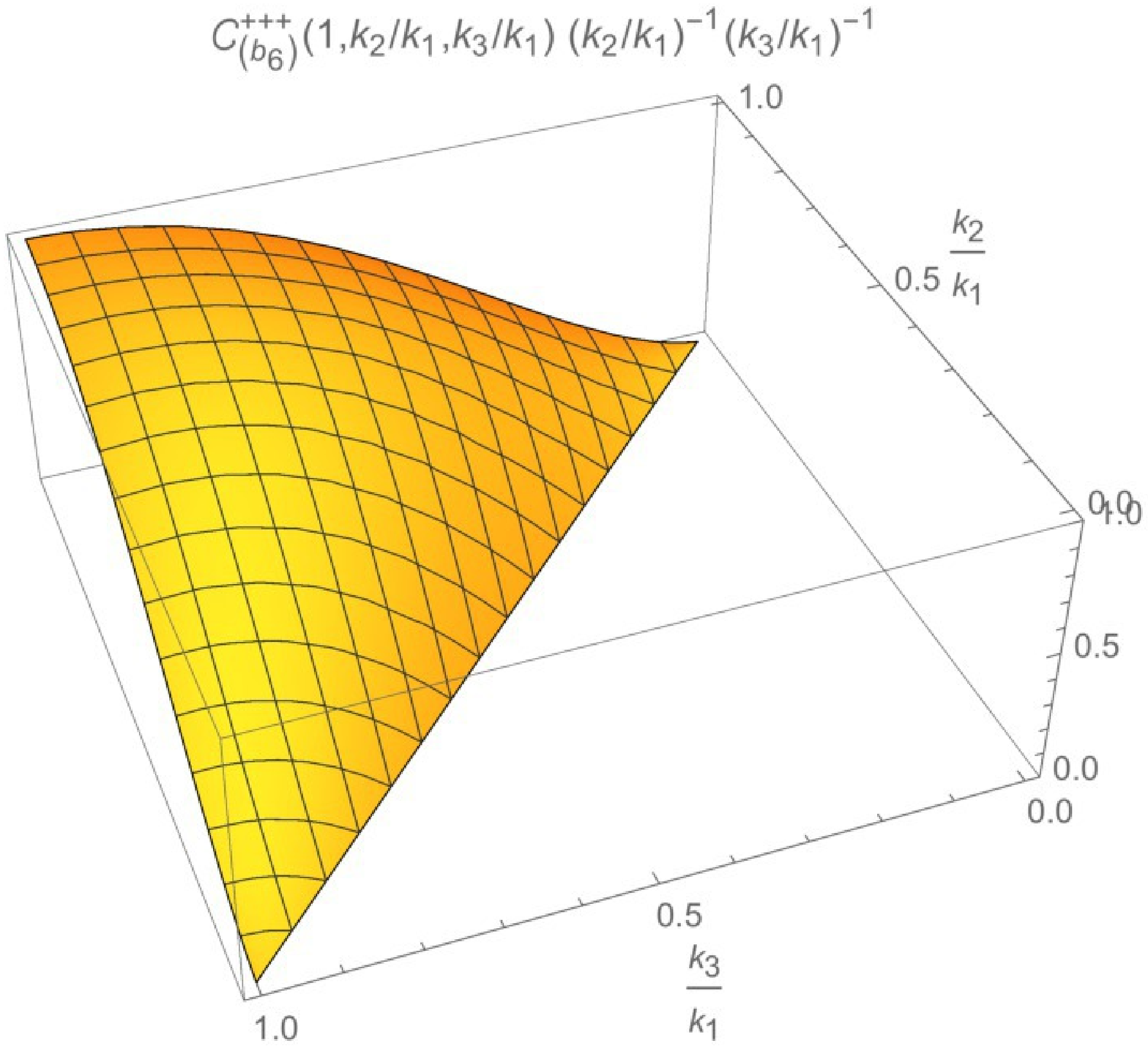}
\caption{$\mathcal C^{+++}_{(b_6)}(1,k_2/k_1,k_3/k_1) (k_2/k_1)^{-1}(k_3/k_1)^{-1}$
as a function of $k_2/k_1$ and $k_3/k_1$. The plot is normalized to 1
for equilateral configurations $k_2/k_1=k_3/k_1=1$.	.}
\label{fig:Cb6+++}
\end{figure}

\begin{figure}[H]
\centering\includegraphics*[width=6.0cm,keepaspectratio,clip]{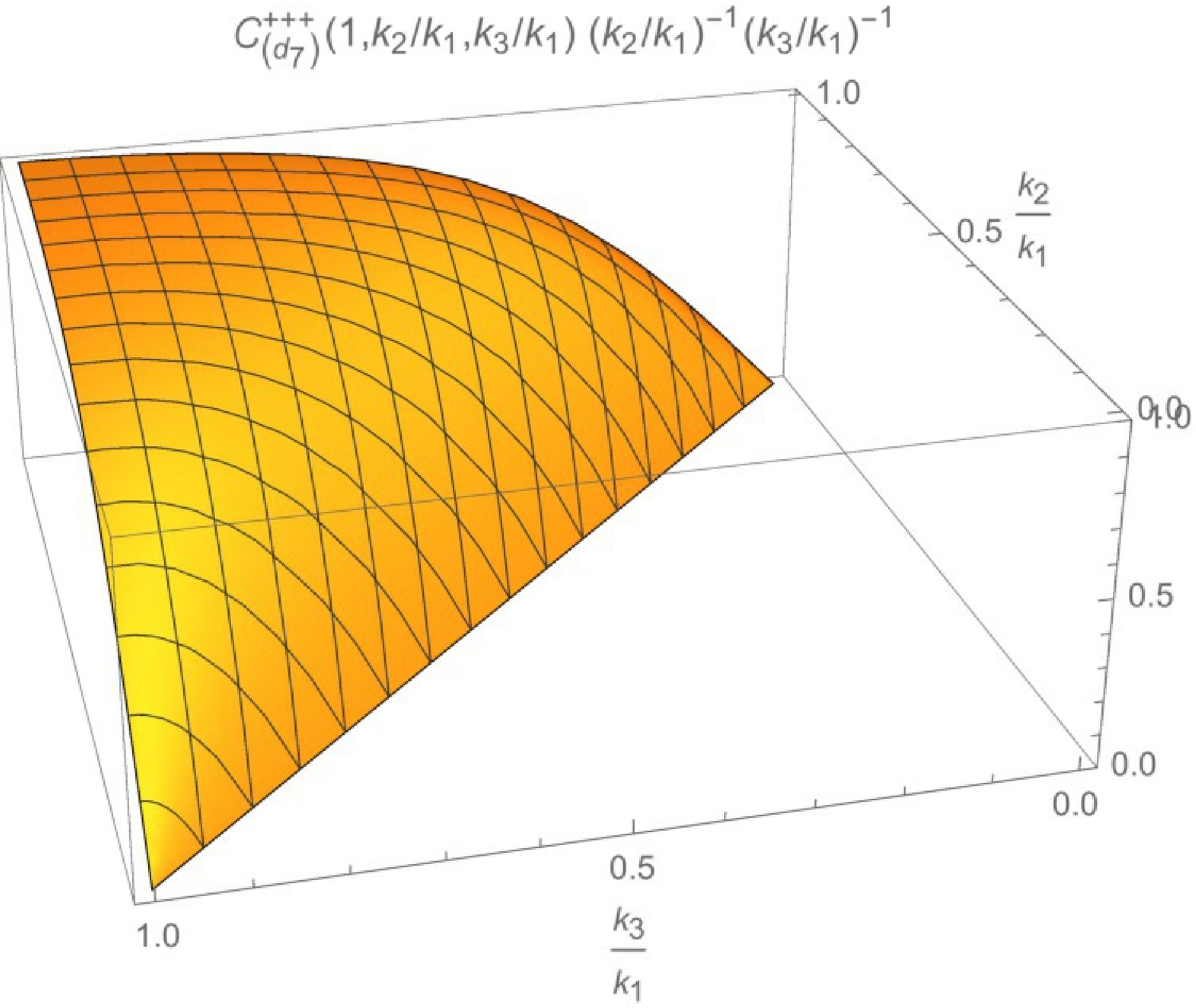}
\caption{$\mathcal C^{+++}_{(d_7)}(1,k_2/k_1,k_3/k_1) (k_2/k_1)^{-1}(k_3/k_1)^{-1}$
as a function of $k_2/k_1$ and $k_3/k_1$. The plot is normalized to 1
for equilateral configurations $k_2/k_1=k_3/k_1=1$.	}
\label{fig:Cd7+++}
\end{figure}

\begin{figure}[H]
\centering\includegraphics*[width=6.0cm,keepaspectratio,clip]{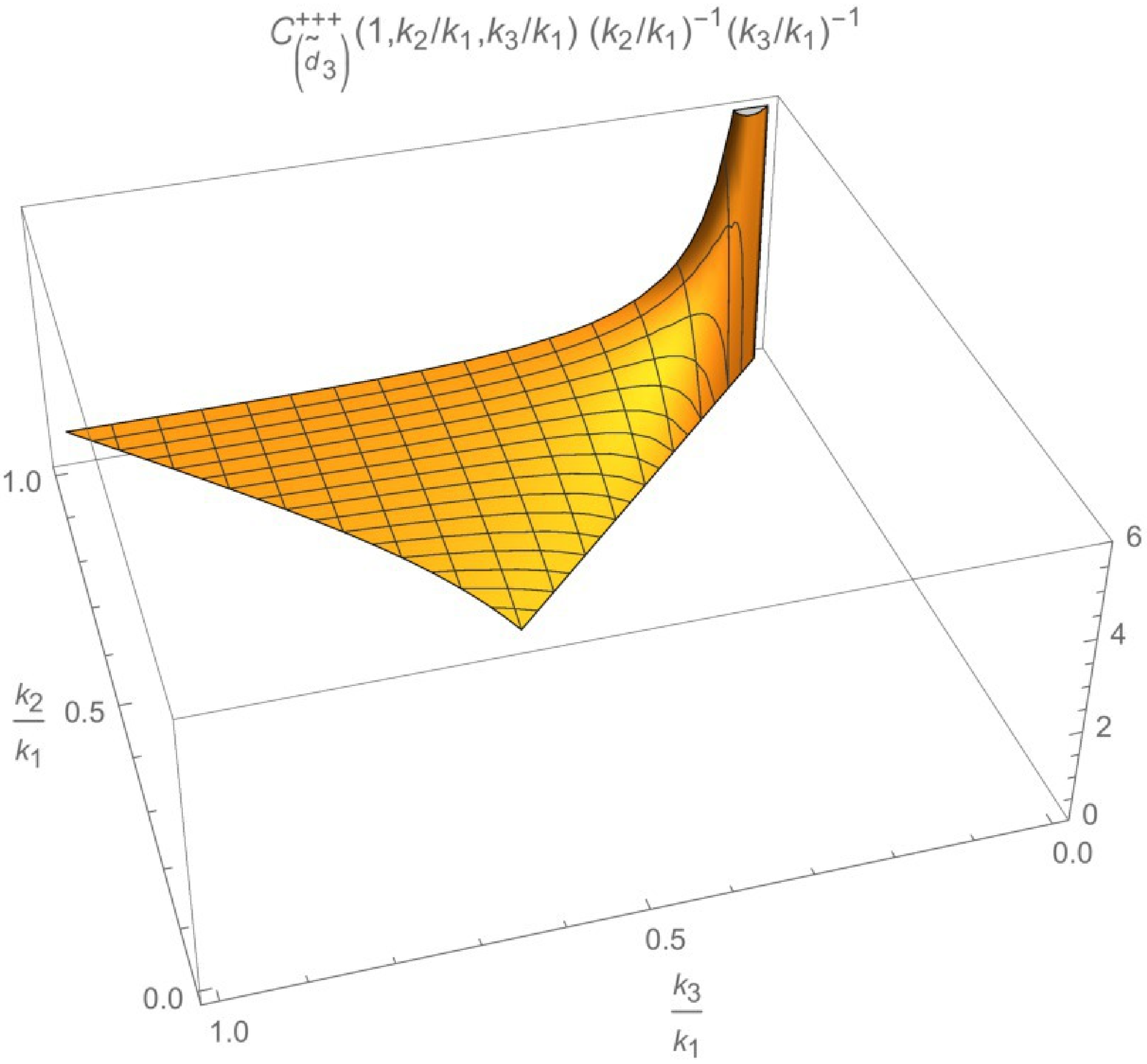}
\caption{$\mathcal C^{+++}_{(\widetilde d_3)}(1,k_2/k_1,k_3/k_1) (k_2/k_1)^{-1}(k_3/k_1)^{-1}$
as a function of $k_2/k_1$ and $k_3/k_1$. The plot is normalized to 1
for equilateral configurations $k_2/k_1=k_3/k_1=1$.	}
\label{fig:Cd3+++}
\end{figure}
\begin{figure}[H]
\centering\includegraphics*[width=6.0cm,keepaspectratio,clip]{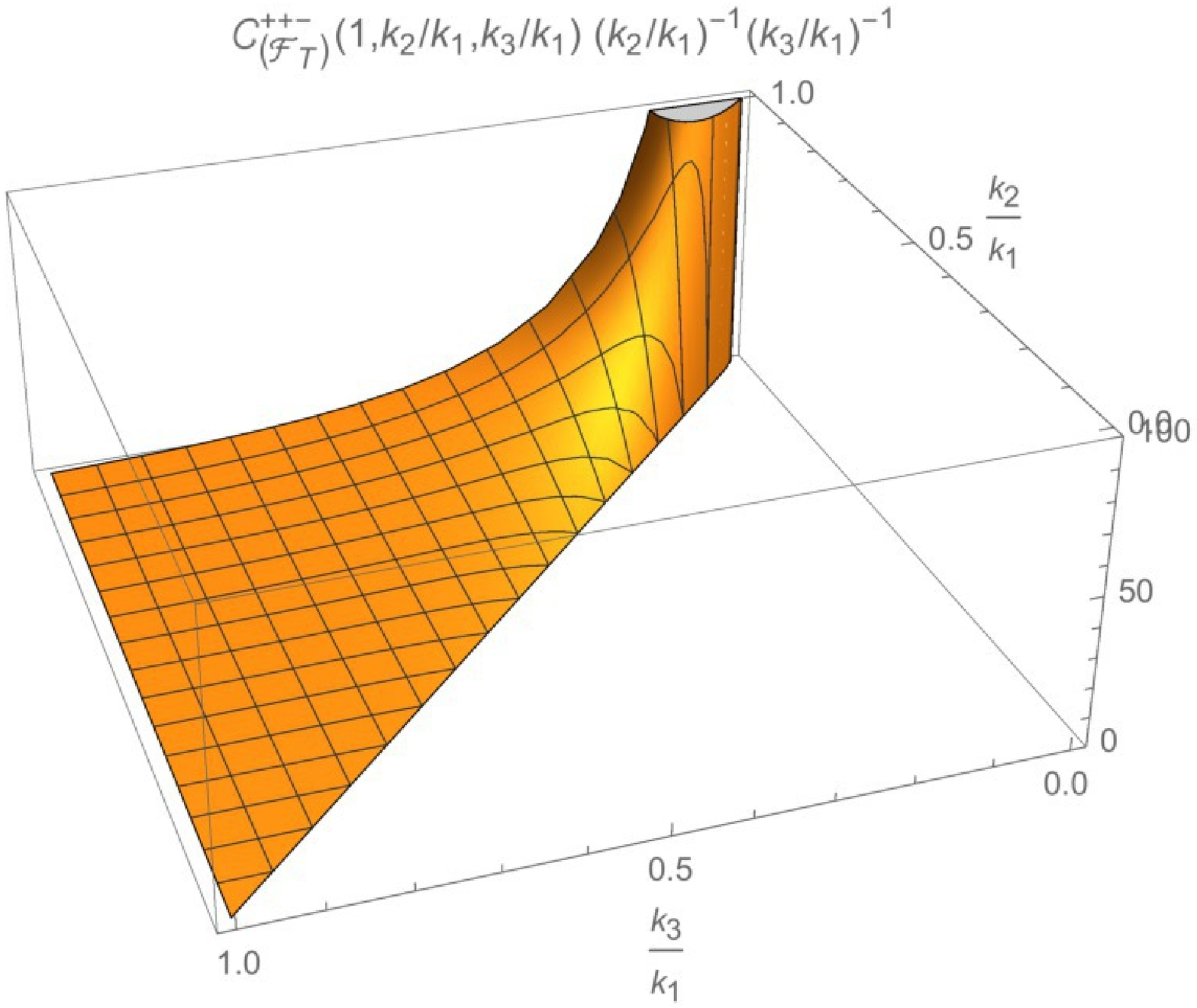}
\caption{$\mathcal C^{++-}_{({\cal F}_T)}(1,k_2/k_1,k_3/k_1) (k_2/k_1)^{-1}(k_3/k_1)^{-1}$
as a function of $k_2/k_1$ and $k_3/k_1$. The plot is normalized to 1
for equilateral configurations $k_2/k_1=k_3/k_1=1$.	}
\label{fig:CGR++-}
\end{figure}
\begin{figure}[H]
\centering\includegraphics*[width=6.0cm,keepaspectratio,clip]{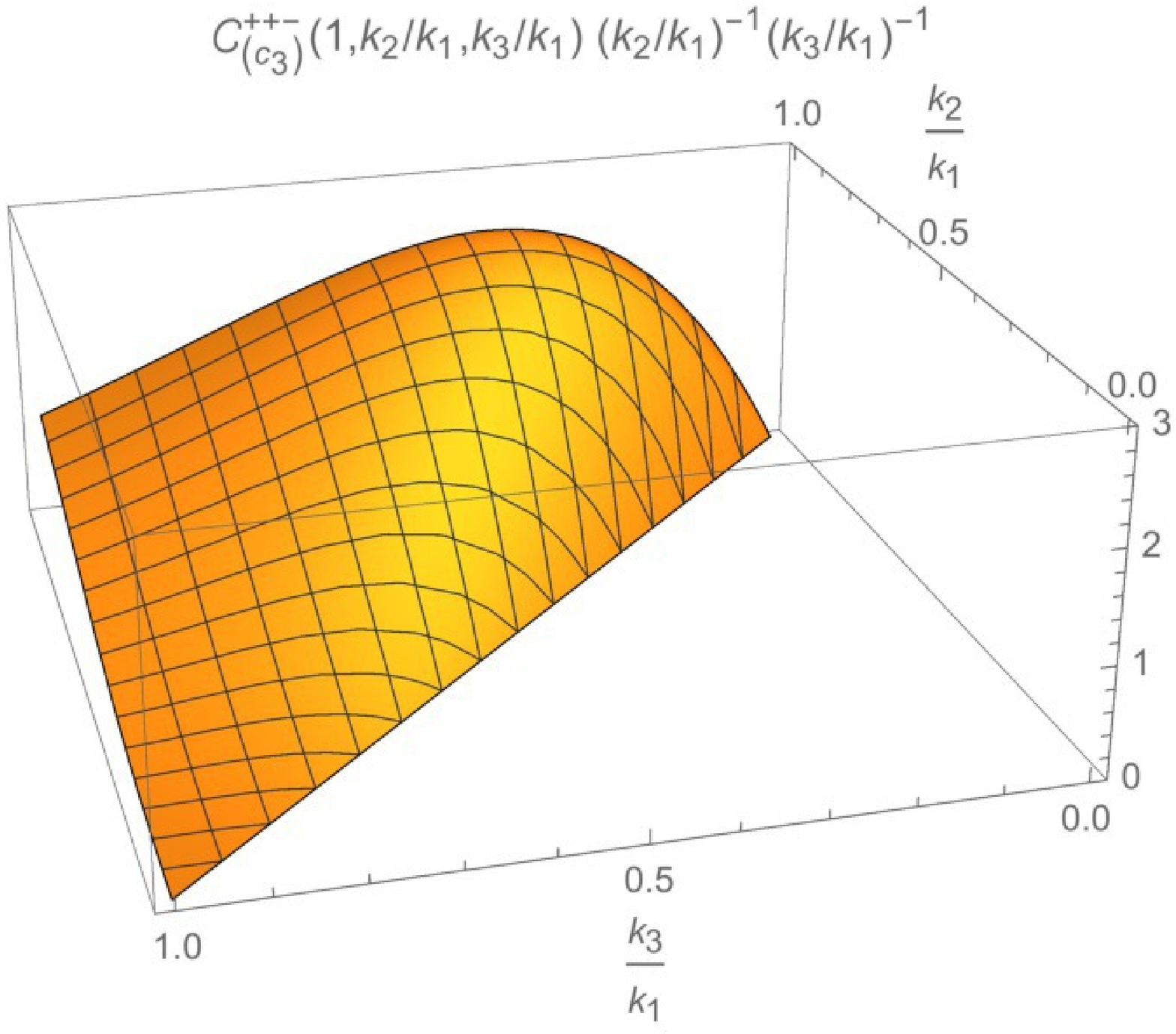}
\caption{$\mathcal C^{++-}_{(c_3)}(1,k_2/k_1,k_3/k_1) (k_2/k_1)^{-1}(k_3/k_1)^{-1}$
as a function of $k_2/k_1$ and $k_3/k_1$. The plot is normalized to 1
for equilateral configurations $k_2/k_1=k_3/k_1=1$.	}
\label{fig:Cc3++-}
\end{figure}

\begin{figure}[H]
\centering\includegraphics*[width=6.0cm,keepaspectratio,clip]{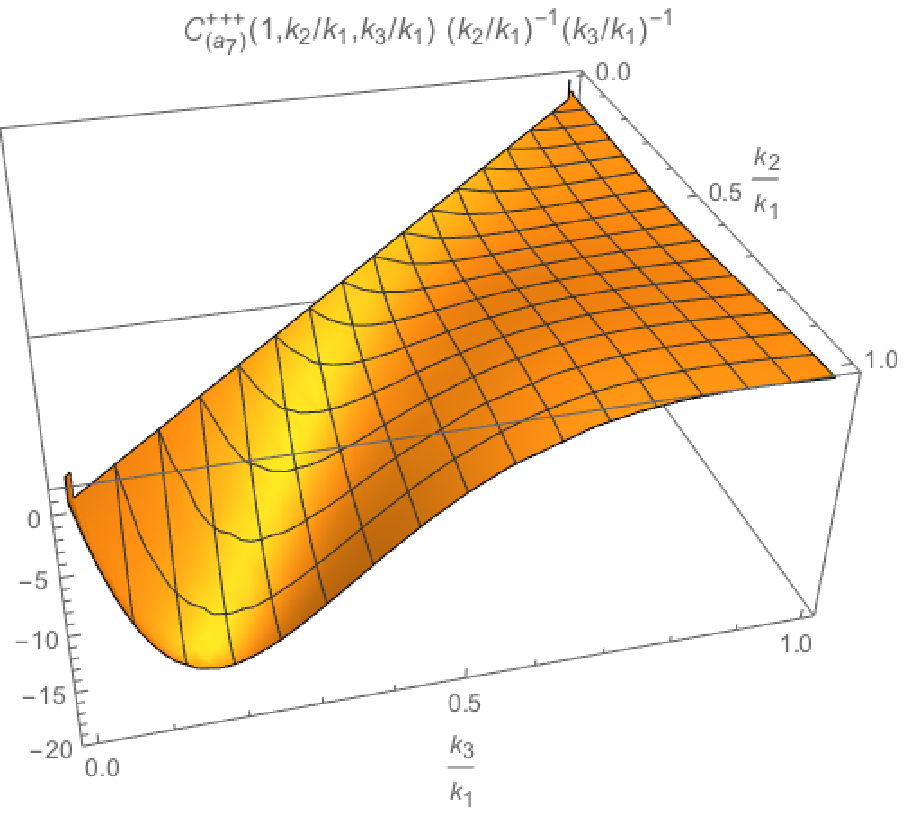}
\caption{$\mathcal C^{++-}_{(a_7)}(1,k_2/k_1,k_3/k_1) (k_2/k_1)^{-1}(k_3/k_1)^{-1}$
as a function of $k_2/k_1$ and $k_3/k_1$. The plot is normalized to 1
for equilateral configurations $k_2/k_1=k_3/k_1=1$.	}
\label{fig:Ca7++-}
\end{figure}
\begin{figure}[H]
\centering\includegraphics*[width=6.0cm,keepaspectratio,clip]{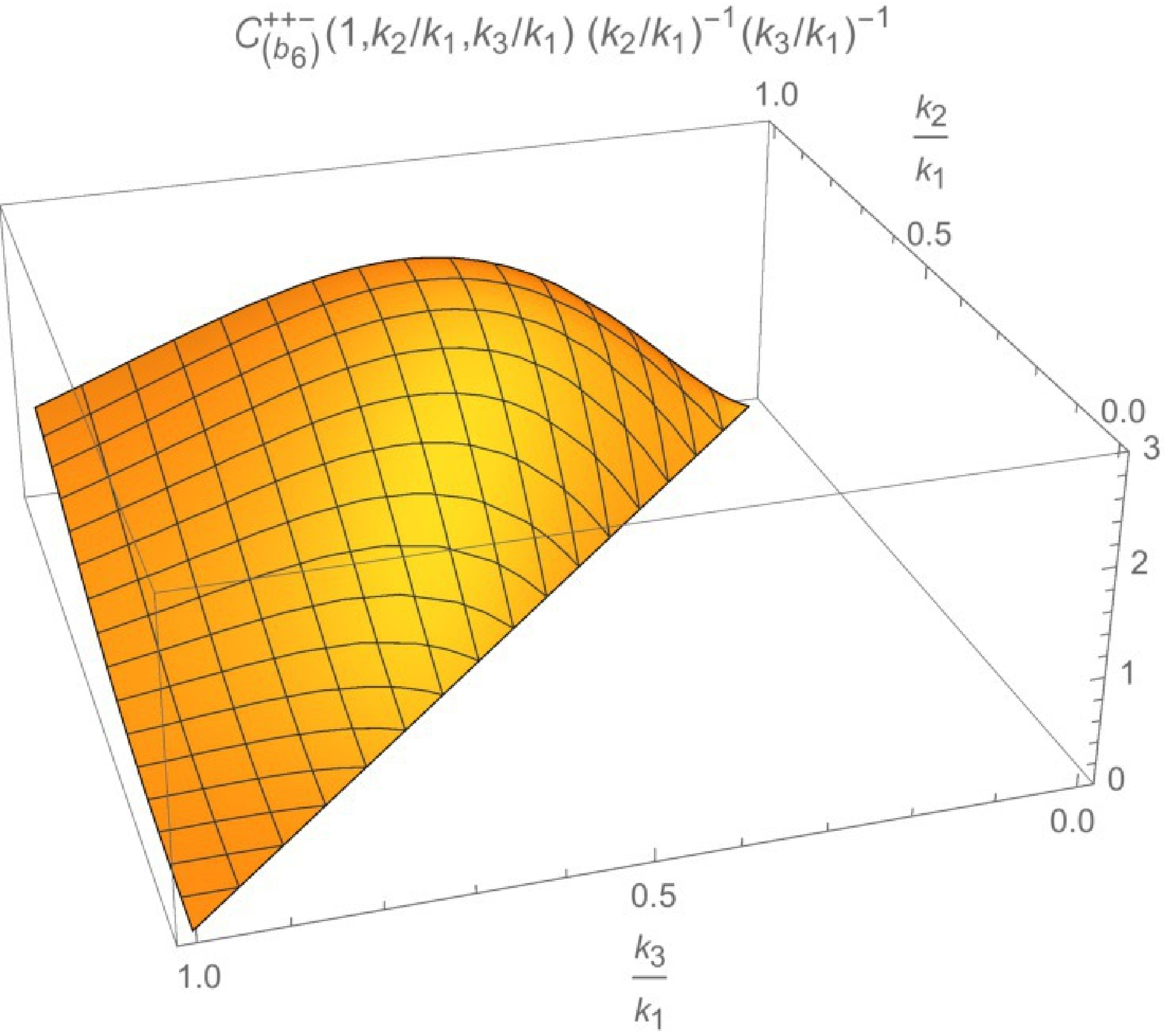}
\caption{$\mathcal C^{++-}_{(b_6)}(1,k_2/k_1,k_3/k_1) (k_2/k_1)^{-1}(k_3/k_1)^{-1}$
as a function of $k_2/k_1$ and $k_3/k_1$. The plot is normalized to 1
for equilateral configurations $k_2/k_1=k_3/k_1=1$.	}
\label{fig:Cb6++-}
\end{figure}

\begin{figure}[H]
\centering\includegraphics*[width=6.0cm,keepaspectratio,clip]{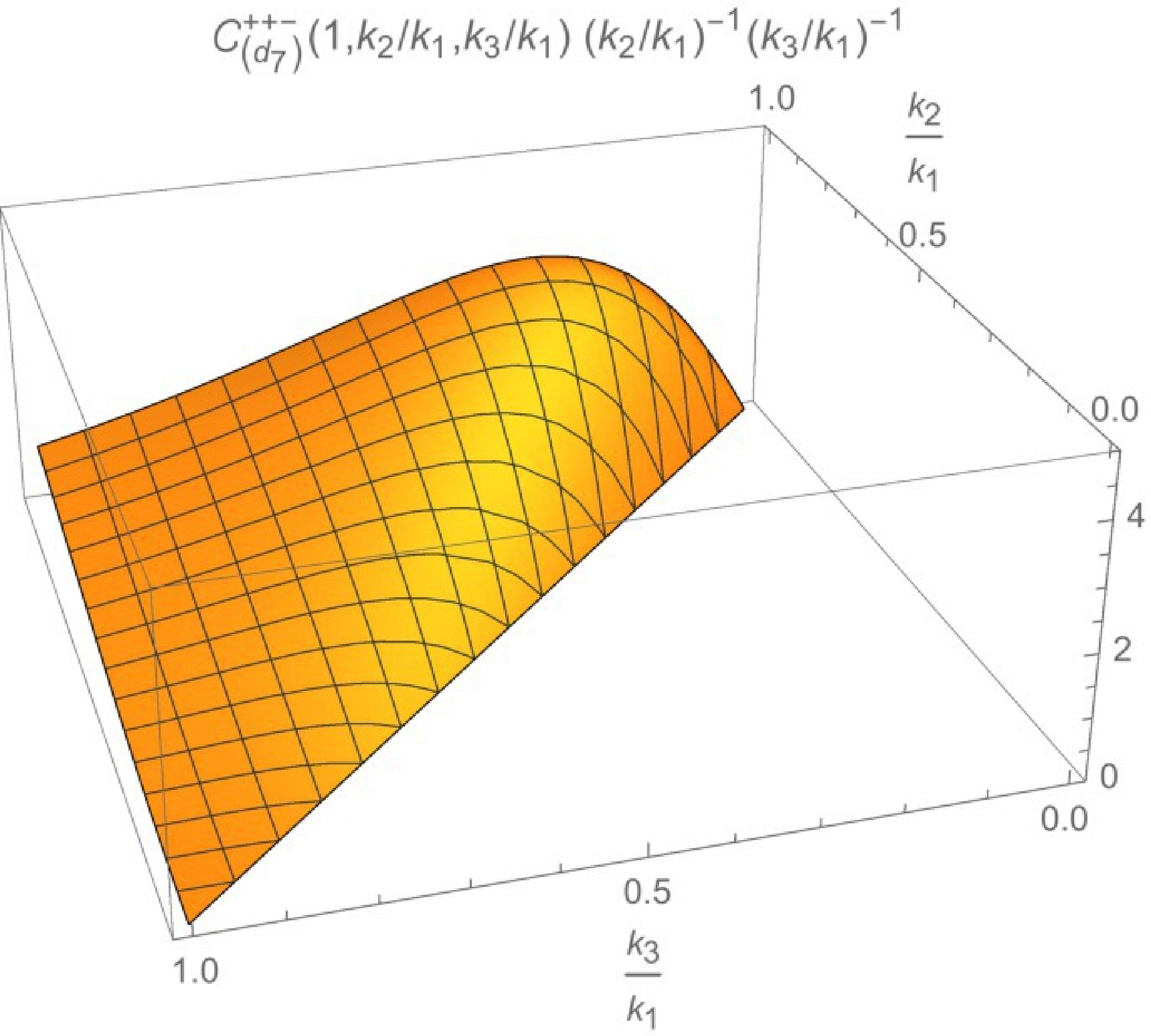}
\caption{$\mathcal C^{++-}_{(d_7)}(1,k_2/k_1,k_3/k_1) (k_2/k_1)^{-1}(k_3/k_1)^{-1}$
as a function of $k_2/k_1$ and $k_3/k_1$. The plot is normalized to 1
for equilateral configurations $k_2/k_1=k_3/k_1=1$.	}
\label{fig:Cd7++-}
\end{figure}
\begin{figure}[H]
\centering\includegraphics*[width=6.0cm,keepaspectratio,clip]{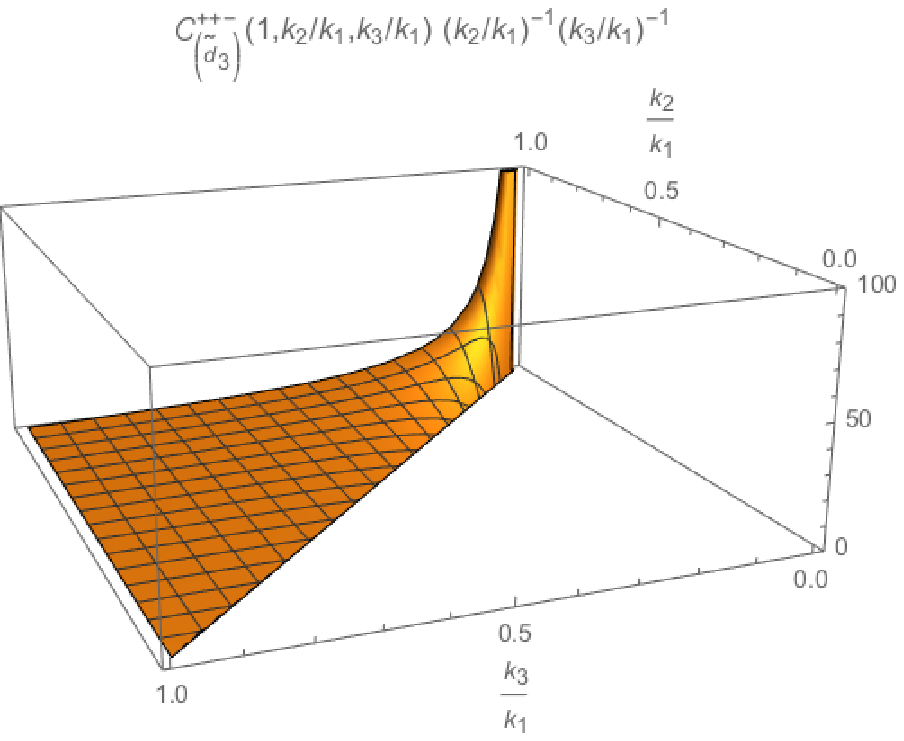}
\caption{$\mathcal C^{++-}_{(\widetilde d_3)}(1,k_2/k_1,k_3/k_1) (k_2/k_1)^{-1}(k_3/k_1)^{-1}$
as a function of $k_2/k_1$ and $k_3/k_1$. The plot is normalized to 1
for equilateral configurations $k_2/k_1=k_3/k_1=1$.	}
\label{fig:Cd3++-}
\end{figure}

The configuration dependences of all the ``$+++$'' and ``$++-$''
amplitudes and the corresponding ${\cal O}(\epsilon^2)$ corrections
are shown in Figs.~\ref{fig:AGR+++}--\ref{fig:Cd3++-}.
At leading order in $\epsilon$ expansion, we find that only
the contribution from ${\cal F}_T$ term, {\em i.e.},
${\cal A}_{({\cal F}_T)}^{+++}$,
peaks in the squeezed limit, while all the other 
${\cal A}_{(\bullet)}^{+++}$'s peak in the equilateral configuration.
We emphasize again that ${\cal A}_{({\cal F}_T)}^{+++}$ {\em is a fixed and universal feature
even for single-field inflation models beyond Horndeski}.
In other words, it is impossible to suppress or enhance ${\cal A}_{({\cal F}_T)}^{+++}$ even in
the present general framework beyond Horndeski as long as the effect of the modified
dispersion relation is small.
It would therefore be of great interest to explore this ``consistency relation''
with CMB B-mode observations such as LiteBIRD~\cite{Matsumura:2013aja}.
When spins are mixed, the momentum dependences of
${\cal A}_{(\bullet)}^{++-}\;(\bullet = c_3, a_7, b_6, d_7)$ are
similar to each other and their peaks
are located between the equilateral and squeezed configurations.
As for the corrections of ${\cal O}(\epsilon^2)$,
it can be seen that they have momentum dependences similar to
their leading order counterparts, namely,
the relative changes in the amplitudes have mild momentum dependences,
as is also the case for the curvature perturbation~\cite{Ashoorioon:2011eg}.
However, ${\cal C}_{(\widetilde d_3)}^{+++}$
shows the peak at a squeezed shape, breaking
the uniqueness of ${\cal A}_{({\cal F}_T)}^{+++}$'s momentum dependence at this order.
Therefore, it would be interesting to investigate the case
where the effect of the modification to the dispersion relation
cannot be treated perturbatively.\footnote{The 3-point function of the
curvature perturbations has been evaluated
when the $k^4$ term is dominant in the context of ghost inflation~\cite{ArkaniHamed:2003uz}.}

\section{Constraints from CMB observations}

We can roughly estimate the constraints on the parameters of the theory by comparing 
the results obtained in the previous section with the Planck results~\cite{Ade:2015ava,Ade:2015lrj}.
Schematically, one has
\begin{eqnarray}
	\frac{\langle hhh\rangle}{\langle hh\rangle\langle hh\rangle}
	\sim \sum\frac{{\cal A}_{(\bullet)}^{s_1s_2s_3}}{k_1 k_2 k_3}
	=\sum\frac{{\cal A}_{(\bullet)}^{s_1s_2s_3}(1,k_2/k_1,k_3/k_1)}{(k_2/k_1)(k_3/k_1)},
	\label{eq:plotted}
\end{eqnarray}
where the quantities in the right hand side are presented in the previous section.
It is reasonable to assume that
\begin{eqnarray}
	 \langle hhh\rangle  \lesssim \langle \mathcal R\mathcal R\mathcal R\rangle,
\end{eqnarray}
where ${\cal R}$ is the curvature perturbation. (Otherwise,
the dominant source of non-Gaussianity in the CMB comes from the tensor modes,
but if so $\langle hhh\rangle$ can be constrained directly from CMB observations.)
Then, using the tensor-to-scalar ratio $r=\langle hh\rangle/\langle {\cal R}{\cal R}\rangle$
and the nonlinearity parameters $f_{\rm NL}^{\cal R}$, we have
\begin{eqnarray}
	\frac{{\mathcal A}_{(\bullet)}^{s_1s_2s_3}}{k_1 k_2 k_3}\lesssim
	\frac{\langle \mathcal R\mathcal R\mathcal R\rangle}{\langle hh\rangle\langle hh\rangle}
	=\frac{1}{r^2}
	\frac{\langle \mathcal R\mathcal R\mathcal R\rangle}
	{\langle\mathcal R\mathcal R \rangle\langle \mathcal R\mathcal R\rangle}
	\sim \frac{f_\mathrm{NL}^{\mathcal R}}{r^2},
\end{eqnarray}
where we have made the second assumption that different ${\mathcal A}_{(\bullet)}^{s_1s_2s_3}$'s
do not cancel each other.
Thus, we obtain the constraint
\begin{eqnarray}
	\frac{{\mathcal A}_{(\bullet)}^{s_1s_2s_3}}{k_1 k_2 k_3}\lesssim \frac{f_\mathrm{NL}^{\mathcal R}}{r^2}
	\sim 10^3 \left(\frac{r}{0.1}\right)^{-2}\left(\frac{f_\mathrm{NL}^{\mathcal R}}{10}\right),	
\end{eqnarray}
which is translated to
\begin{align}
\frac{c_3H }{{\cal G}_T},\;
\frac{a_7H^3}{{\cal G}_Tc_h^4},\;
\frac{b_6H^2}{{\cal G}_Tc_h^2},\;
\frac{d_7H^4}{{\cal G}_Tc_h^6}
\lesssim
10^3 \left(\frac{r}{0.1}\right)^{-2}\left(\frac{f_\mathrm{NL}^{\mathcal R}}{10}\right),
\end{align}
or, equivalently,
\begin{align}
\frac{c_3c_h}{H},\; \frac{a_7H}{c_h^3},\;\frac{b_6}{c_h},\;\frac{d_7H^2}{c_h^5}
\lesssim 
10^{13} \left(\frac{r}{0.1}\right)^{-3}\left(\frac{{\cal P}_{\cal R}}{10^{-9}}\right)^{-1}
\left(\frac{f_\mathrm{NL}^{\mathcal R}}{10}\right),
\end{align}
where ${\cal P}_{\cal R}$ is the power spectrum of the curvature perturbation.
Note the mass dimensions of the coefficients: $[c_3]=1,~[b_6]=0,~[a_7]=-1,$ and $[d_7]=-2$.
Thus, the current constraints are very weak.

\section{Discussions and Conclusions}

In this paper, we have clarified primordial non-Gaussianities of tensor modes
by computing the tensor 3-point functions within the unifying framework of
scalar-tensor theories~\cite{Gao:2014soa}.
The framework includes the Horndeski theory and recent GLPV generalization
as specific cases.
We have shown that no new terms appear in the cubic Lagrangian
even if one goes to the GLPV theory beyond Horndeski.
In more general theories beyond GLPV, we have found four new interactions, one of which
is related to the modification of the dispersion relation in the linear theory.
The impact of this modification of the dispersion relation can be parametrized
using a small parameter $\epsilon$,
and we have computed the 3-point functions analytically by
treating $\epsilon$ as a small expansion parameter.
Two of the four new terms beyond GLPV have already been studied
to leading order in $\epsilon$ in the context of Ho\v{r}ava gravity~\cite{Huang:2013epa},
and our results are in agreement with those in Ref.~\cite{Huang:2013epa} where they overlap.
In Ref.~\cite{Gao:2011vs} it was found that
there are only two independent terms in the cubic Lagrangian within Horndeski:
the standard one present already in general relativity
generating squeezed non-Gaussianity with the {\em fixed} amplitude
and the other nonstandard one predicting equilateral non-Gaussianity. 
We have strengthened this statement by showing that
at leading order in the $\epsilon$ expansion
the squeezed non-Gaussianity is only generated by
this ``standard'' cubic term and has the fixed amplitude, ${\cal A}_{({\cal F}_T)}^{+++}$,
even in the general unifying framework of scalar-tensor theories.
At leading order in $\epsilon$ all the other non-Gaussian amplitudes
peak at equilateral shapes.
The ``standard'' interaction is quite likely to be present
in the cubic Lagrangian because it can easily be generated from
the term linear in the Ricci scalar.
The fixed and universal nature of ${\cal A}_{({\cal F}_T)}^{+++}$
thus provides us the ``consistency relation'' in the primordial tensor sector.
Any detection of the equilateral tensor non-Gaussianity
would imply the nonstandard interactions between gravity and the scalar degree of freedom,
though it is difficult to distinguish among different contributions.
Note that the inverse is not true;
even in the absence of equilateral non-Gaussianity,
gravity could be modified significantly from general relativity,
because one can consider various nonstandard interactions that
do not affect the tensor sector, as well as $R_i^j\delta K_j^i$.

We have found that the effects of the modified dispersion relation
appear in the non-Gaussian amplitudes at ${\cal O}(\epsilon^2)$.
The momentum dependences of the correction $\mathcal{C}_{(\bullet)}^{s_1s_2s_3}$ 
are similar to their leading order counterparts $\mathcal{A}_{(\bullet)}^{s_1s_2s_3}$.
It should be noted that the correction to the squeezed non-Gaussianity,
$\mathcal{C}_{({\cal F}_T)}^{s_1s_2s_3}$, breaks its fixed and universal nature.
Therefore, detection of squeezed tensor non-Gaussianity whose amplitude
is different from ${\cal A}_{({\cal F}_T)}^{+++}$ would
imply a significant higher order term in the dispersion relation
or the tensor modes of non-inflationary origin.

Let us mention here some other sources of gravitational waves from the early universe,
as it would be interesting to explore tensor non-Gaussian signatures of such origin
in order to contrast them with the results in this paper.
A spectator scalar field other than inflaton can produce gravitational waves~\cite{Biagetti:2013kwa},
though it is difficult to expect large amplitudes~\cite{Biagetti:2014asa,Fujita:2014oba}.
Large tensor modes are produced, {\em e.g.}, by
self-ordering of multi-component scalar fields after a global phase
transition~\cite{JonesSmith:2007ne,Fenu:2009qf,Giblin:2011yh,Figueroa:2012kw,Kuroyanagi:2015esa}.
Vector fields can also source gravitational waves during inflation~\cite{Mukohyama:2014gba,Barnaby:2010vf}.
We hope that we will come back to the issue of
non-Gaussian signatures of those gravitational waves in future publications.

Finally, it should be noted that by construction the present analysis
does not cover multi-field inflation models. Though the multi-field effects
in the tensor sector are expected to be small, it would be interesting
to investigate whether one can discriminate single- and multi-field models
using primordial non-Gaussianity of tensor modes.

\acknowledgments 
This work was supported in part by the JSPS Grant-in-Aid for Scientific
Research Nos.~24740161 (T.K.) and
MEXT KAKENHI No.~15H05888 (T.K.).


\appendix

\section{Disformal transformation}

Let us investigate the transformation properties of
the quadratic and cubic actions under
the disformal transformation,
\begin{align}
\hat g_{\mu\nu}=A(\phi,X)g_{\mu\nu}+B(\phi,X)\partial_\mu\phi\partial_\nu\phi.
\end{align}
It is more convenient to write this in the unitary gauge as
\begin{eqnarray}
\hat g_{\mu\nu}=e^{2\alpha}\left[g_{\mu\nu}+\left(1-e^{2\beta}\right)n_\mu n_\nu\right],
\end{eqnarray}
where $n_\mu$ is the unit normal to the constant time hypersurfaces, $n_\mu=(-N, 0)$.
Since we are interested in the tensor perturbations on a cosmological background,
we may only consider the case where both $\alpha$ and $\beta$
are functions of time and do not fluctuate.
Under the above disformal transformation, the ADM variables transform as
\begin{align}
\hat\gamma_{ij}=e^{2\alpha}\gamma_{ij},\quad
\hat N=e^{\alpha+\beta}N,\quad \hat N^i=N^i,
\end{align}
yielding
\begin{align}
\hat K_i^{j}&=e^{-\alpha-\beta}\left(K_i^j+\frac{\dot\alpha}{N}\delta_i^j\right),
\\
\hat R_i^j&=e^{-2\alpha}R_i^j.
\end{align}
Substituting these to the Lagrangian~(\ref{eq:relevant action}),
we find that the disformal transformation maintains the form of the Lagrangian
for the tensor perturbations:
\begin{eqnarray}
	\frac{{\cal L}}{\sqrt{-g}}\to \frac{{\cal L}}{\sqrt{-\hat g}}
	&=&
	\hat d_1 \hat R+\hat d_3 \hat R_i^j\hat R_j^i + \hat d_7 \hat R_i^j\hat R_j^k\hat R_k^i
	+\hat b_2\hat{\delta K}_i^j\hat{\delta K}_j^i +\hat c_3\hat{\delta K}_i^j\hat{\delta K}_j^k\hat{\delta K}_k^i
		\nonumber\\&&
	+\hat a_2\hat R_i^j\hat{\delta K}_j^i
	+\hat a_7\hat R_i^j\hat R_j^k\hat{\delta K}_k^i
	+\hat b_6\hat R_i^j\hat{\delta K}_j^k\hat{\delta K}_k^i, 
\end{eqnarray}
where
\begin{align}
&\hat d_1=e^{-2\alpha-\beta}\widetilde d_1,
\quad
\hat d_3=e^{-\beta}\widetilde d_3,
\quad
\hat d_7=e^{2\alpha-\beta}d_7,
\quad
\hat b_2=e^{-2\alpha+\beta}b_2,
\notag\\
&\hat c_3=e^{-\alpha+2\beta}c_3,
\quad
\hat a_2=e^{-\alpha}\widetilde a_2,
\quad
\hat a_7=e^{\alpha}a_7,
\quad
\hat b_6=e^\beta b_6,
\end{align}
and we dropped the terms that are irrelevant to the tensor perturbations.
This, in particular, implies that
\begin{eqnarray}
{\cal G}_T\to\hat {\cal G}_T=e^{-2\alpha+\beta}{\cal G}_T,
\quad
{\cal F}_T\to\hat {\cal F}_T=e^{-2\alpha-\beta}{\cal F}_T.
\end{eqnarray}
Using the two time-dependent functions $\alpha$ and $\beta$,
one can fit ${\cal G}_T$ and ${\cal F}_T$ to the standard form, {\em i.e.},
$\hat{\cal G}_T=\hat{\cal F}_T=\mpl^2$~\cite{Creminelli:2014wna}.
However, the higher order term in the dispersion relation
cannot be removed~\cite{Fujita:2015ymn}.
It is also clear that one cannot make further simplifications
in the cubic action by the use of the disformal transformation.


\end{document}